\documentclass[aps,prl,twocolumn,floatfix,superscriptaddress]{revtex4}  % for review and submission
\usepackage{graphicx}  % neededfor figures
\usepackage{bm}        % for math
\usepackage{amssymb}   % for math
\usepackage{amsmath}
\usepackage{hyperref}
\usepackage[latin1]{inputenc}

\usepackage{color}

\newcommand{\ud}[1]{{#1^{\dagger}}}
\newcommand{\bra}[1]{\left\langle #1\right|}
\newcommand{\ket}[1]{\left| #1\right\rangle}

\newcommand{\mean}[1]{\langle #1\rangle}

\begin{document}
\title{Subwavelength vacuum lattices and atom-atom interactions in photonic crystals}

 \author{A. Gonz\'{a}lez-Tudela}
 \affiliation{Max-Planck-Institut f\"{u}r Quantenoptik Hans-Kopfermann-Str. 1.
85748 Garching, Germany }

 \author{C.-L. Hung}
 \affiliation{Norman Bridge Laboratory of Physics 12-33}
  \affiliation{Institute for Quantum Information and Matter, California Institute of Technology, Pasadena, CA 91125, USA}
      
\author{D. E. Chang}
 \affiliation{ICFO-Institut de Ciencies Fotoniques, Mediterranean Technology Park, 08860 Castelldefels (Barcelona), Spain} 

\author{J. I. Cirac}
 \affiliation{Max-Planck-Institut f\"{u}r Quantenoptik Hans-Kopfermann-Str. 1.
85748 Garching, Germany }

 \author{H. J. Kimble}
  \affiliation{Max-Planck-Institut f\"{u}r Quantenoptik Hans-Kopfermann-Str. 1.
85748 Garching, Germany }
 \affiliation{Norman Bridge Laboratory of Physics 12-33}
  \affiliation{Institute for Quantum Information and Matter, California Institute of Technology, Pasadena, CA 91125, USA}

\date{\today}

\begin{abstract}
We propose the use of photonic crystal structures to design subwavelength optical lattices in two dimensions for ultracold atoms by using both Guided Modes and Casimir-Polder forces. We further show how to use Guided Modes for photon-induced large and strongly long-range interactions between trapped atoms. Finally, we analyze the prospects of this scheme to implement spin models for quantum simulation.
\end{abstract}

% \pacs{42.50.Nn, 73.20.Mf, 71.36.+c}S
\maketitle

Quantum simulation with cold atoms in optical lattices~\cite{bloch12a} constitutes an attractive avenue for the exploration of quantum many-body physics~\cite{cirac12a}. One of the main challenges in the field is to increase the energy and length scales involved in current setups, as this would potentially reduce both the temperature and coherence-time requirements and introduce new long-range physics. In this Letter we present a new paradigm for high-density, two-dimensional ($2$-D) optical lattices in photonic crystal waveguides (PCWs) \cite{joannopoulos11a}. We show that specially engineered two-dimensional photonic crystals provide a practical platform to both trap atoms and engineer their interactions, in a way that surpasses the limitations of current technologies and enables the exploration of new forms of quantum many-body matter. Our schemes remove the constraint on lattice constant set by the free-space optical wavelength in favor of deeply sub-wavelength atomic arrays with lattice constant 
$\simeq 50$nm. We further describe new possibilities for atom-atom interactions mediated by photons in PCWs with energy scales several orders of magnitude larger than by way of exchange interactions in free-space lattices and with the capability to engineer strongly long-range interactions.

%%%%
Obstacles to the exploration of quantum many-body physics with cold atoms ~\cite{bloch12a,cirac12a} are the small energy scales ($\sim 10^3-10^4$Hz) and the restriction to nearest neighbor interactions in free-space optical lattices. Alternate approaches for quantum simulation include dipolar molecules \cite{micheli06a} and Rydberg atoms \cite{jaksch00a,lukin01a}, which give rise to weakly long-range interactions typically scaling as $1/r^\alpha$, where $\alpha=3$ and $r$ is the distance between the atoms (using standard notation \cite{bouchet10a} $\alpha>$ dimensionality). 
Other possibilities involve lattices with period below that for a free-space optical standing wave by using plasmonic \cite{gullans12a} or superconducting systems \cite{romeroisart13a}.

\begin{figure*}[!t]
 \centering
 \includegraphics[width=0.95\linewidth]{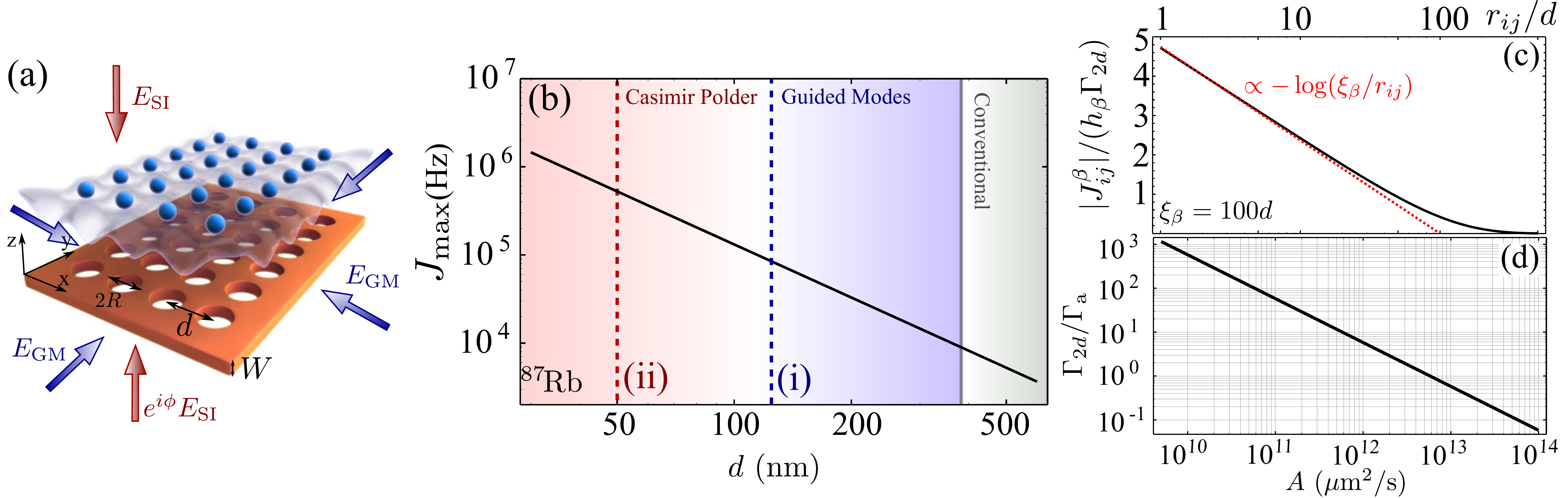}
 \caption{(a) General scheme for a nanophotonic lattice: dielectric slab of thickness $W$ and refractive index $n$ with a lattice of holes of radius $R$ and periodicity $d$. Optical trapping is by way of guided modes (GM) and side illumination (SI). (b) Scaling of maximum tunneling [$J$ in Eq. 1] as a function of lattice periodicity $d$ for Rb. The vertical dashed lines correspond to the examples explained in the manuscript. (c) Spatial dependence of spin-spin interaction, $J_{ij}^\beta/(h_\beta \Gamma_{2d})$, as a function of distance $r_{ij}/d$ for a situation where the atomic frequency lies in the bandgap and $\xi_\beta=100 d$. (d) Scaling of $\Gamma_{2d}/\Gamma_{\mathrm{a}}$ with guided mode band curvature $A$ using the geometrical parameters of the TE mode of Figs. \ref{fig3}-\ref{fig4} within the isotropic approximation.}
 \label{fig1}
  \vspace{-4mm}
 \end{figure*}

In this Letter, we show that the integration of ultra-cold atomic physics with nano-photonics opens up new avenues for the creation of quantum many-body matter. As illustrated in Fig.~\ref{fig1}(a), $2$-D arrays of atom traps can be generated by optical-dipole forces from `Guided Modes' (GMs) of PCWs whose refractive index $n(\mathbf{r})$ is modulated with a period $d<\lambda_0/2$, where $\lambda_0$ is the vacuum wavelength \cite{joannopoulos11a}. Atoms can also be trapped in $2$-D `vacuum lattices' arising from the spatial variation of Casimir-Polder (CP) forces \cite{buhmann07a} near a PCW. Such atom lattices with $d<\lambda_0/2$ yield larger energy scales for quantum simulation than is generally possible with conventional free-space optical lattices.

We further show that PCWs provide versatile means for creating atom-atom interactions mediated by photons within the GMs of the PCW. These effective atomic interactions can be very large and \emph{strongly} long-range. By operating with a Raman transition either within a band gap or in a dispersive regime for the PCW, the dynamics of atom-atom interactions can be predominantly conservative or dissipative, with the possibility to make this choice in real time.
%%%%

Our analyses are based upon recent experiments for atom localization near nanoscopic dielectrics \cite{vetsch10a,goban12a,thompson13a,goban13a,yu14a,tiecke14a} and related theoretical proposals (e.g., self-organization \cite{chang13a} and coherent atom-atom interactions \cite{shahmoon13a,douglas13a}). By advancing atomic lattices with PCWs from one to two dimensions, we gain access to a richer set of phenomena, including Hubbard physics with large interaction energies, quantum magnetism with the capability to `design' the strength and range of the interaction, and topological phases for $2$-D atom lattices with photon mediated interactions \cite{cho08a,diehl11a,bardyn13a}.

Because of the wide range of physical phenomena, our Letter is structured to present first an overview of the parameter space that is opened by our proposals. We then describe particular examples that illuminate our strategies and analysis techniques. Details of the physics underlying these calculations are given in the accompanying Supplementary Material \cite{SupMat}.

\paragraph{Hubbard physics in optical lattices with $d < \lambda_0/2$.}

Conventional investigations employ free-space optical traps with lattice constant $d=\lambda_0/2$. By moving to 2-D planar PCWs as in  Fig.~\ref{fig1}(a), lattices with $d<\lambda_0/2$ become possible for quantum simulation with both fermonic and bosonic atoms. For definiteness, we consider bosonic atoms, for which the following Bose-Hubbard hamiltonian has been well established ($\hbar=1$) \cite{jaksch98a}:

\vspace{-4mm}
\begin{equation}
\label{bosehub}
H_{\mathrm{BH}}=-J\sum_{\mean{i,j}}\ud{b_i} b_j+U\sum_{i} n_i (n_i-1)/2\,,
\end{equation}
\vspace{-3mm}

\noindent where  $b_i$ ($\ud{b_i}$) annihilates (creates) a localized atom on site $\mathbf{r}_i$ and $n_i=\ud{b_i} b_i$. The tunneling rate $J$ and on-site interaction energy $U$ are both upper bounded by the so-called \emph{recoil} energy of the lattice, $E_R= h^2/8 m d^2$.

Figure~\ref{fig1}(b) displays results for the scaling of the maximum tunelling rate $J_{\mathrm{max}}$ for $^{87}$Rb atoms trapped in lattices with $50$nm $\lesssim d \lesssim 300$nm as compared to $\lambda_0/2 \sim 385$nm. With only optical confinement in the GM region of Fig.~\ref{fig1}(b), the upper limit for $J,U$ could be increased by about $10\times$ relative to a free-space lattice. Further reductions in $d$ are enabled by a novel trap design that uses vacuum forces (CP) for lateral confinement \cite{yannopapas08a,contrerasreyes10a} in the $x,y$ plane and optical forces along $z$ perpendicular to a planar PCW, leading to the possibility for a $60$-fold increase in $J,U$ for $^{87}$Rb.

\paragraph{Strong spin-spin interactions mediated by photons.}

Another exciting perspective for quantum simulation is the realization of spin-spin hamiltonians for quantum magnetism of the general form

\vspace{-4mm}
\begin{equation}
 \label{genspin}
 H_{\mathrm{spin}}=\sum_{\beta=x,y,z}\sum_{i,j}J_{ij}^\beta \sigma_i^{\beta} \sigma_j^\beta\,,
\end{equation}
\vspace{-3mm}

\noindent where $\sigma_{i}^{\beta}$ are Pauli operators and $J_{ij}^\beta$ are the spin-spin interaction energies in the $\beta$ direction for sites ${i,j}$. In free-space lattices, $|J_{ij}^\beta|/2\pi \lesssim 10^3$~Hz~\cite{trotzky08a} and the range of interactions is restricted to nearest neighbors. To extend the interaction range, dipolar molecules~\cite{micheli06a} and Rydberg atoms~\cite{jaksch00a,lukin01a} yield spatial decay of $1/r_{ij}^3$. By employing GMs for photon-mediated atomic interactions, we demonstrate schemes for simulating spin-models as in Eq.~\ref{genspin} with atoms trapped within 2-D PCWs. The dynamics can be either conservative or dissipative by varying the effective detuning,  $\Delta_{\beta}$, between laser and band-edge frequencies. For the conservative regime with the atomic transition within a band gap and $\Delta_{\beta}<0$,
\begin{equation}
 \label{J}
  J^{\beta}_{ij}=h_{\beta} \Gamma_{2d} K_0 (|r_{ij}|/\xi_\beta)\,,
\end{equation}
\vspace{-4mm}

\noindent where $K_0 (|r_{ij}|/\xi_\beta)$ is a modified Bessel function of the second kind that dictates the spatial dependence of $J^{\beta}_{ij}$. This function decays as $K_0(x)\propto \log (1/x)$, when $x \ll 1$ and $K_0(x)\propto e^{-x}/\sqrt{x}$ when $x \gg 1$,  with $\xi_\beta=\sqrt{|A/\Delta_\beta|}$ a tunable length [Fig.~\ref{fig1}(c)], which depends both on the curvature of the bands, $A$, and effective detuning $\Delta_\beta$. Real-time variation of $\Delta_\beta$ can be used to switch between the different scaling regimes. The interaction strength is $h_{\beta} \Gamma_{2d}$, where $\Gamma_{2d}$ describes the atom-guided mode coupling [Fig.~\ref{fig1}(d)] and $h_{\beta}$ is fixed by the Rabi frequency and detuning of an external laser in our two-photon Raman coupling scheme (see below). 

As illustrated in Fig.~\ref{fig1}(d), $\Gamma_{2d}/2\pi \sim 10^6-10^9$ Hz (using $\Gamma_\mathrm{a}/2\pi= 6.07$~MHz for the line D2 of Rb in free space), depending on the geometrical and material characteristics of the PCWs, e.g.,  curvature $A$ of 
the band. The projected values for $J^{\beta}_{ij}$ are several 
orders of magnitude larger than with state-of-the-art methods in free-space lattices. The stronger and longer range of interactions favour frustrations and foresees the observation of more stable supersolid phases \cite{buchler07a,hauke10c,maik12a}, long-lived metastable states \cite{trefzger08a} or the `instantaneous' transmission of correlations after local quenches \cite{hauke13a}.

By contrast, operating outside the bandgap in a dispersive regime leads to dissipative evolution described by a master equation. In this case, our proposal leads to simulations of \emph{strongly} long-range dissipative interactions, and thereby opens new opportunities for the dissipative generation of entanglement \cite{verstraete09a,gullans12a,gonzaleztudela13a} and steady-state topological phases \cite{diehl11a,bardyn13a} for $2$-D atom lattices.

\paragraph{Optical lattices with planar PCWs.}

The regime labeled `Guided Modes' (GMs) in Fig.~\ref{fig1}(b) can be studied by considering a dielectric slab with a square lattice of circular holes of periodicity $d$ as in Fig.~\ref{fig1}(a). The evanescent fields of counter-propagating GMs along $x$ and $y$ create a periodic far-off resonance optical trapping (FORT) potential proportional to $|\Omega_{\mathrm{GM}}(\mathbf{r})|^2/\delta$, where $\delta$ is the detuning between the GM and atomic frequencies and $|\Omega_{\mathrm{GM}}(\mathbf{r})|^2$ is proportional to the field density $|\mathbf{E}_{\mathrm{GM}}(\mathbf{r})|^2$. For definiteness, we analyze the lowest order transverse magnetic (TM) GMs with polarizations predominantly along $z$ for $k_x,k_y$. To stabilize the lattice in the vertical direction, a third pair of lasers counter propagates along $z$ (side illumination, SI) with wavelength $\lambda_{\mathrm{SI}}$ and Rabi frequency $\Omega_{\mathrm{SI}}$. These $z$-beams have the same frequency but amplitudes related by a $e^{i\phi}$, thereby 
localizing a minimum near the surface of the PCW \cite{thompson13a}. The vertical trapping minima, $z_t(x,y)$, must be near the surface of the PCW to form an $\{x,y\}$ lattice in the rapidly decaying fields of the GMs (e.g., $z_t = 65$~nm for the dashed line (i) in Fig.~\ref{fig1}(b) corresponding to a situation with $d=125$~nm). In such close proximity to the dielectric, surface forces from Casimir-Polder (CP) interactions must be taken into account, which we calculate following the procedures in Refs. \cite{buhmann07a,rodriguez09a,hung13a}. In \cite{SupMat} we provide a detailed analysis of the trapping potentials for the example of line (i) in Fig.~\ref{fig1}(b).

Optimization of the lattice potential formed from GMs, SI, and CP is straightforward for periodicity $\lambda_0/2n < d < \lambda_0/2$, but becomes problematic for $d \lesssim \lambda_0/2n$. First of all note that in the limit of a uniform slab (i.e., $n(\mathbf{r}) \rightarrow n$), the lower bound for the periodicity an optical lattice with GMs is $d_{\mathrm{min}}= \lambda_0/2 n$, obtained for a thick slab. Hence, $d_{\mathrm{min}}$ is reduced by a factor $n$ relative to $d=\lambda_0/2$ for free-space lattices. Reductions in $d$ below $d_{\mathrm{min}}$ are possible for PCWs optical lattices. However, PCWs rapidly lose contrast for the GM intensity and the resulting FORT potential $V_{\mathrm{GM}}(x,y)$ as $d$ decreases below $d_{\mathrm{min}}$. Moreover, decreasing $d$ brings a requirement for trapping at $z_t$ closer to the surface and a concomitant increase in CP forces. These two factors result in a steep rise in laser intensities required for stable trapping~\cite{SupMat}.

%%%%%%%%%%%%%%%%%%%%%%%%%%%%%%%%%%%

\begin{figure}[!tb]
 \centering
 \includegraphics[width=0.95\linewidth]{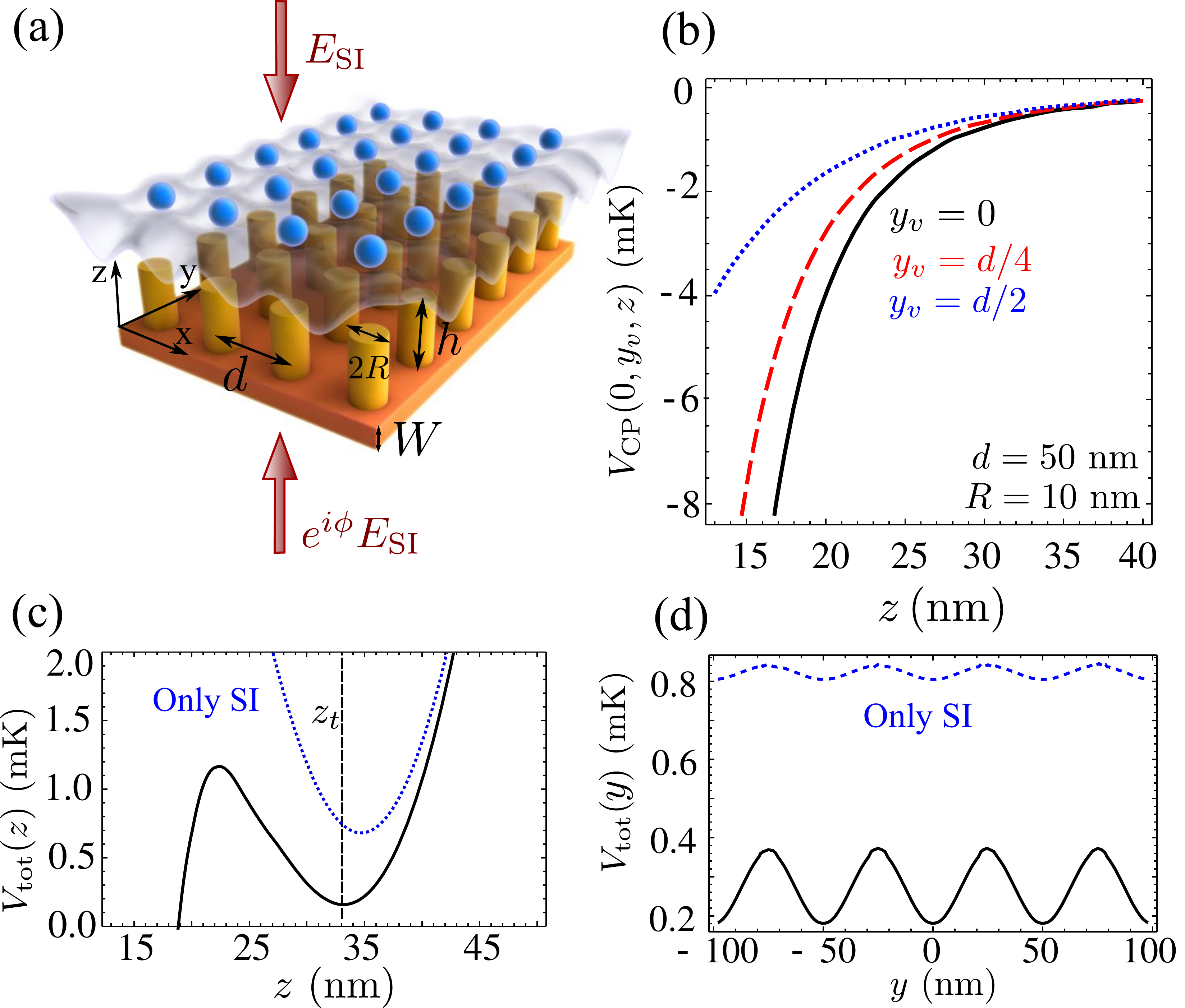}
 \caption{(a): Schematic of a dielectric slab of thickness $W$ and refractive index $n_{\mathrm{subs}}$ with dielectric posts of height $h$ with refractive index $n_2$. For illustration we use GaP for post and substrate structure and treat $^{87}$Rb. (b): CP potential $V_{\mathrm{CP}}(0,y,z)$ vertical cut for three different horizontal positions in the unit cell: $y_v=0$ (solid black), $y_v=d/2$ (dashed red) and $y_v=d$ (dotted blue) for a structure with $d=50$ nm and $R=10$ nm. (c): Vertical cut of the total (solid black) and SI potential (dotted blue) above posts surface. Chirping the SI phase $\phi$ moves the $z_t$ minimum and lattice depth. (d) Horizontal cut of the  total potential (solid black) and SI potential (dotted blue) at $z_t=32.5$ nm. Both panels (c-d) are for GaP with $d=50$ nm, $R=0.2d$ and $W=h=118.75$ nm. Vertical trapping is provided by SI with $\lambda_{\mathrm{SI}}=760$ nm,  $\Omega_{\mathrm{SI}}/2\pi = 130$~GHz for $E_{\mathrm{SI}}$ and $\phi=1.7$~radians.
 Horizontal trapping is provided by CP modulation.}
 \label{fig1b}
  \vspace{-4mm}
 \end{figure}
 
\paragraph{2-D lattices with vacuum forces.}

To create lattices with $d < d_{\mathrm{min}}$, we introduce a new method for trapping that uses CP interactions as a tool rather than a hinderance for localization near a surface. In the spirit of Ref.~\cite{hung13a}, we exploit that a periodic modulation of refractive index $n(x,y)$ creates a periodic $x,y$ modulation of the CP potential $V_{\mathrm{CP}}(x,y,z)$ \cite{yannopapas08a,contrerasreyes10a}. Large transverse wave vectors $k_x,k_y$ associated with $V_{\mathrm{CP}}(x,y,z_t)$ for $k_0 z_t \ll 1$ avoid fundamental constraints on contrast for $V_{\mathrm{GM}}(x,y,z_t)$ in subwavelength structures. 

A proof-of-principle example of our scheme is shown schematically in Fig.~\ref{fig1b}(a) and consists of a periodic array of cylindrical posts in a deeply sub-wavelength regime with $d=50$ nm. The lattice for trapping in the $x,y$ plane is predominantly due to the CP potential $V_{\mathrm{CP}}(x,y,z)$, while trapping along $z$ is via $V_{\mathrm{SI}}(x,y,z)$. We compute $V_{\mathrm{CP}}(x,y,z)$ numerically \cite{buhmann07a,rodriguez09a,hung13a} and in Fig.~\ref{fig1b}(b) display vertical cuts of $V_{\mathrm{CP}}(0,y_v,z)$ at $y_v=\{0, d/2, d\}$ showing the $z$-dependence of $V_{\mathrm{CP}}(0,y_v,z)$ as the planar position moves from the axis of a rod at $\{x,y\}=0$. Although, we focus on a unit cell, a 2-D array of posts is included in our calculation. 
The total trapping potential for the atomic lattice is $V_{\mathrm{tot}}(x,y,z)=V_{\mathrm{CP}}(x,y,z)+V_{\mathrm{SI}}(x,y,z)$, with line cuts shown in Fig.~\ref{fig1b}(c-d). For Fig.~\ref{fig1b}(c), the trap minima are chosen to be at a vertical distance $z_t\approx 32.5$ nm to achieve the required contrast in the $x$-$y$ plane for Bose-Hubbard physics of Eq.~\ref{bosehub}. Figure \ref{fig1b}(d) clearly demonstrates that the dominant contribution to the variation of $V_{\mathrm{tot}}(x,y,z_t)$ in the $x$-$y$ plane is the `vacuum-lattice' from CP interactions and not $V_{\mathrm{SI}}(x,y,z_t)$.

Trap depth of the 2D vacuum lattice can be dynamically tuned over a wide range by adjusting the vertical trap position $z_t$ \cite{SupMat}. The trap depth $V_d$ and frequencies $\omega_{\mathrm{t}}$ for Fig.~\ref{fig1b} are $\{V_{d,\mathrm{xy}},V_{d,\mathrm{z}}\}/2\pi \approx \{3.5,20.8\}$~MHz and $\{\omega_{t,\mathrm{xy}},\omega_{t,\mathrm{z}}\}/2\pi \approx \{1.7,4.2\}$~MHz. In the $x,y$ plane, the trapping depth of Fig.\ref{fig1b} is $\sim 15 E_R$, which guarantees the possibility of having localized Wannier modes in the lattice \cite{greiner03a}. If we used only SI FORT potential, the trap depth would be $\sim 3 E_R$, which does not lead to localization in a unit cell. The associated scattering rate of the trap scales as $V_{\mathrm{SI}} \Gamma_{\mathrm{a}}/\delta$, yielding to scattering rates  $\sim 2\pi\times 10$~Hz using $\Gamma_{\mathrm{a}}$ of Rb. The results in Fig.~\ref{fig1b}, corresponding to Fig.~\ref{fig1}(b)(ii), and further analyses \cite{SupMat} suggest 
that `vacuum lattices' could provide significant increases in the energy scale for Bose-Hubbard physics with cold atoms.

\paragraph{Long-range interactions mediated by photons in PCWs.}

Our discussion thus far relates to the CP and GM regimes in Fig.~\ref{fig1}(b) with an emphasis on reduced lattice constant relative to free-space. However, PCWs also offer exciting opportunities for investigations of many-body physics via photon-mediated interactions among atoms trapped within a PCW \cite{john90a}. Although $1$-D PCWs have been principally considered \cite{shen05a,shahmoon13a,douglas13a,hung13a,goban13a,yu14a}, our schemes for $2$-D PCWs yield physics beyond the $1$-D case. Moreover, we provide full descriptions of the trapping configuration, band design and photon-mediated couplings using realistic parameters.

\begin{figure}[tb]
  \centering
  \includegraphics[width=0.9\columnwidth]{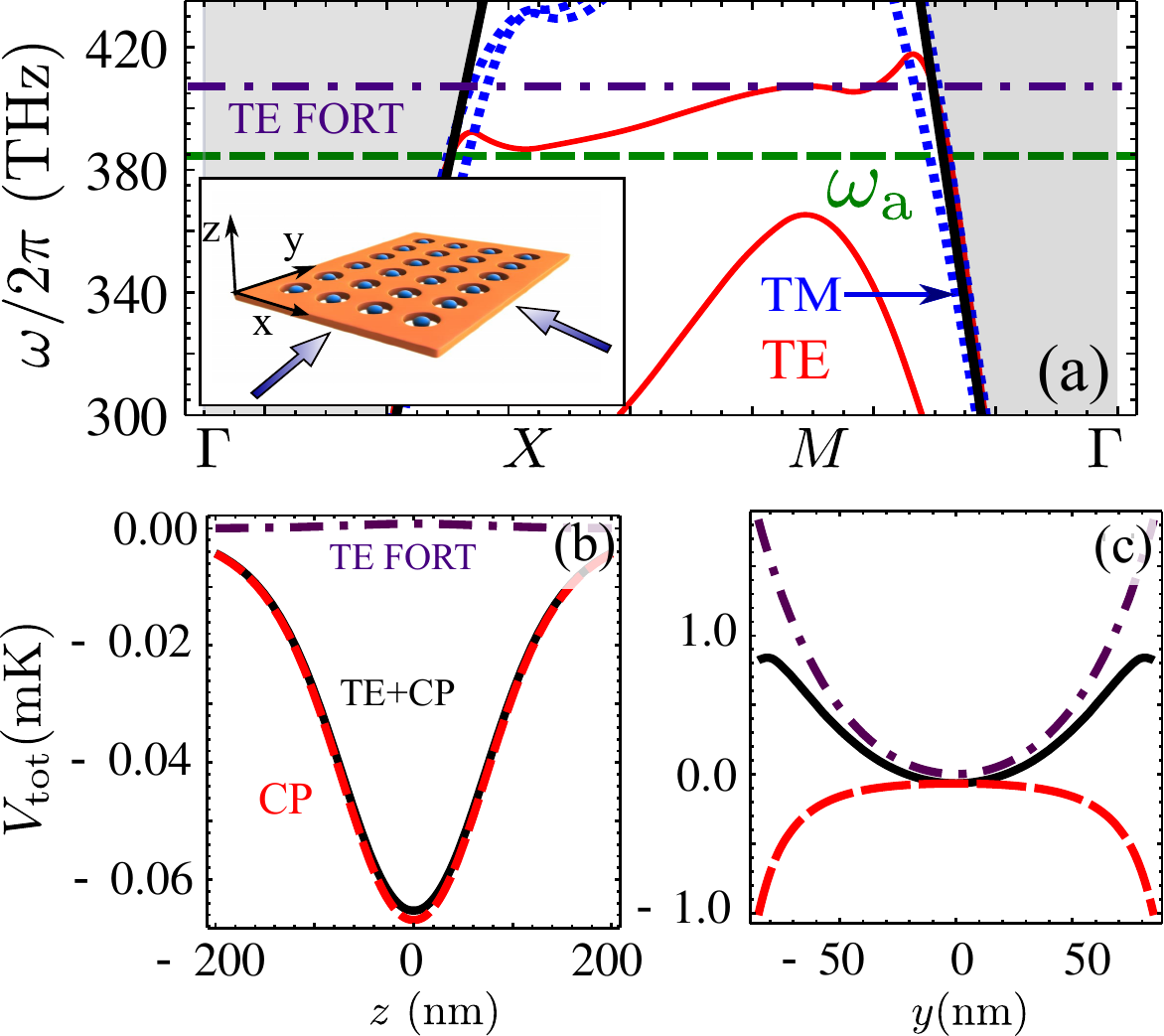}
  \vspace{-3mm}
  \caption{(a) Band structure of a GaP photonic crystal slab with a square lattice of holes with  $W=0.316 d$, $R=0.4d$ and $d=316$ nm. Dashed green line (dashed-dotted purple) denotes atomic frequency (GM frequency for trapping). Thick black line represents the light-line above which the modes are non-guided. Inset: General scheme for atom trapping inside holes of the structure. (b) Total potential (black) and its different contributions: TE FORT (dashed-dot purple) and CP (dashed red) as a function of the vertical distance $z$, with origin $z=0$ in the center of the hole. (c) Same as in (b) but for a horizontal cut in $y$ at the center of the hole $z=x=0$. $\lambda_0$ and $\Omega_{\mathrm{GM}}/2\pi$ are approximately $730$ nm and $35$~GHz.}
  \vspace{-5mm}
  \label{fig3}
  \end{figure}
  
Figure \ref{fig3}(a) displays the band structure for a square lattice of holes in a GaP slab \cite{johnson01a}. Our interest is in the guided modes shown in the area below the light line, while the shaded area above the light line represents the continuum of leaky modes. This structure supports a bandgap for the transverse electric (TE-like) guided modes (in red). To achieve photon mediated atomic interactions, we design the band structure to provide a band of guided modes suitable for atom trapping and for large atom-field interactions. Off-resonant excitation with TE-guided modes along $\{k_x,k_y\}$ near the $M$-point of the band diagram creates a FORT lattice in the $x$-$y$ plane that compensates the CP forces in these directions. Atoms are trapped at $z_t=0$ using vertical confinement provided by CP forces \cite{hung13a}. By further including two SI beams counter propagating along $z$, it is also possible to control the position of the minima in $z$ and load the trap. Line cuts of the total trap 
potential $V_{\mathrm{tot}}$ are given in Fig.~\ref{fig3}(b,c), with the contributions from the FORT (purple) and CP (red) potentials shown. The trap depth $V_d$ and frequencies $\omega_{\mathrm{t}}$ for the particular example in Fig.~\ref{fig3} are $\{V_{d,\mathrm{xy}},V_{d,\mathrm{z}}\}/2\pi \approx \{18.7, 1.3\}$MHz and $\{\omega_{t,\mathrm{xy}},\omega_{t,\mathrm{z}}\}/2\pi \approx  \{0.67,0.16\}$MHz.

\begin{figure}[tb]
  \centering
 \includegraphics[width=0.8\linewidth]{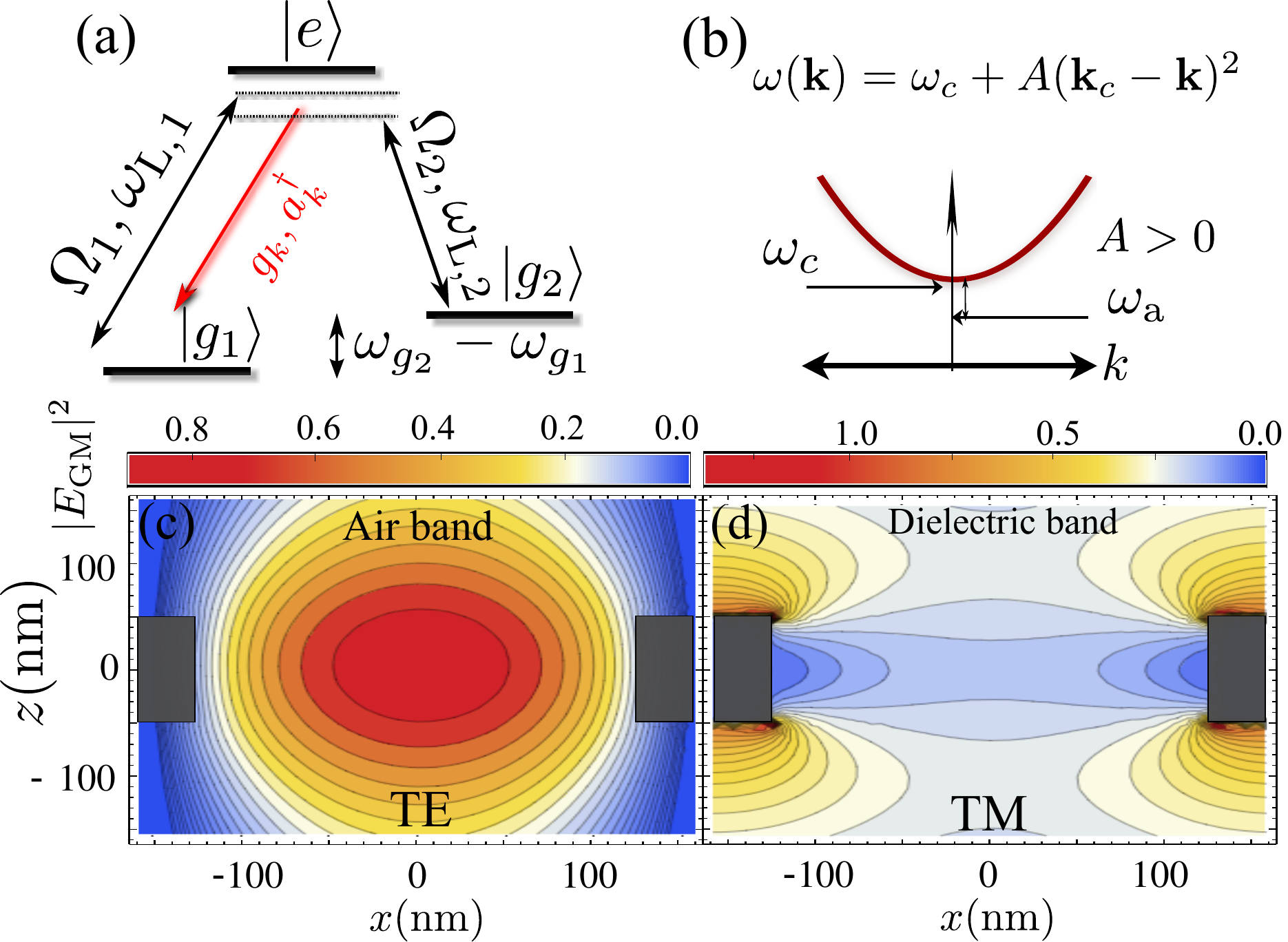}
   \vspace{-4mm}
  \caption{(a) Lambda configuration to engineer the coupling to a single polarization of the PCW continuum. (b) Parabolic approximation of the band structure close to a band edge. (c) [(d)] $|E_{\mathrm{GM}}(x,0,z)|^2$ for the TE [TM] polarized GMs of the air [dielectric] band with frequency around $\omega_{\mathrm{a}}$ of a GaP PCW slab with a square lattice of holes with $d=316$ nm, $R=0.4d$, $W=0.316 d$. Dark shaded areas represent the regions of dielectric material.}
  \vspace{-4mm}
  \label{fig4}
  \end{figure}

The Rb atoms trapped in the centers of the holes in Fig.~\ref{fig3}(b,c) interact with TE GMs near the band edge at the $X$ point with GM frequencies approaching that of Rb $D_2$ line. As shown in Fig.~\ref{fig4}(a) we consider an atomic $\Lambda$ scheme driven by two off-resonant lasers with detunings $\Delta_l=(\omega_e-\omega_{g_l})-\omega_{L,l}$ for $l=1,2$. The polarizations are chosen such that the transition $g_1 \leftrightarrow e$ interacts with the TE guided modes near the $X$-point (which are polarized predominantly along $y,x$ for $k_x,k_y$). $g_2 \leftrightarrow e$ interacts instead with transverse magnetic (TM-like) modes (which are polarized principally along the $z$-direction). Mode profile intensities are shown in Figs.~\ref{fig4}(c-d), with large (small) coupling strength at the trap site for the TE (TM) GMs with energies close to $\omega_{\mathrm{a}}=\omega_e-\omega_{g_1}$, which lies within the band gap. 

Assuming the coupling with the GMs can be treated perturbatively and in the limit where $|\Delta_l|\gg \Omega_l$ ($l=1,2$) for the scheme of Fig.~\ref{fig4}(a), both the excited states and the GMs can be adiabatically eliminated leading to interactions betweeen the two-state spins $\{\ket{g_1}_i,\ket{g_2}_i\}$, which results in an effective master equation \cite{breuerbook02a}:
\vspace{-1mm}
\begin{align}
\label{meq}
 \dot{\rho}=\sum_{i,j} \sum_{\beta=xy,z} \Gamma_{ij}^{\beta}\big( \mathcal{O}^\beta_i\rho (\mathcal{O}^{\beta}_j)^\dagger-(\mathcal{O}^\beta_j)^\dagger \mathcal{O}^\beta_i\rho\big)+\mathrm{h.c} \Big]\,, 
\end{align}
\vspace{-3mm}

\noindent where $\Gamma^{\beta}_{ij}=\gamma^\beta_{ij}/2+i J^\beta_{ij}=h_{\beta} \Gamma_{2d} \mathrm{F}^\beta(r_{ij})$ and $\mathcal{O}^{\beta}_i=\ket{g_1}_i\bra{g_2}\,, \sigma_i^z$ for  $\beta=xy\,, z$, respectively. Here $h_\beta=\big(\Omega_l/(2\Delta_{\beta})\big)^2$ with $l=1\,(2)$ for $\beta=z\, (xy)$. $\mathrm{F}^\beta(r_{ij})$ is a function whose form depends on whether $\{\Delta_{xy}, \Delta_z\}  \lessgtr 0$, where $\Delta_{xy}=\omega_{g,2}-\omega_{g,1}+\omega_{L,2}-\omega_c$ and $\Delta_z=\omega_{L,1}-\omega_c $ \cite{SupMat}.  The imaginary contribution of the collective coupling, $J^\beta_{ij}$, accounts for the coherent evolution and the real part, $\gamma^\beta_{ij}$, describes collective dissipation.

To find approximate expressions for these spin-spin interactions, we use a parabolic approximation of $\omega(\mathbf{k})$ as depicted in Fig.~\ref{fig4}(b) and assume that around the $X$ point both the coupling and $\omega(\mathbf{k})$ behave isotropically. We have also performed numerical integration with the exact energy dispersion and coupling of the structure obtaining similar scalings \cite{SupMat}. When $\Delta_{\beta}>0$, $\mathrm{F}^\beta(r_{ij})$ has both real and imaginary components given by

\vspace{-0mm}
\begin{equation}
 \Gamma^\beta_{ij}|_{\Delta_{\beta}>0}= \frac{\pi}{2}h_{\beta} \Gamma_{2d} H^{(1)}_0\big[|r_{ij}|/\xi_\beta\big]\,,
\end{equation}
where $H^{(1)}_0(x)=J_0(x)+i Y_0(x)$ is a Hankel function of the first kind, and the length scale $\xi_\beta=\sqrt{|A/\Delta_{\beta}|}$ determines the range and strength of the correlations that can be controlled independently via the detuning, $\Delta_{\beta}$ \cite{douglas13a}. The Bessel function $J_0(x)\propto 1$ for $x \ll 1$ and $\propto 1/\sqrt{x}$ for $x \gg 1$.

By contrast, when $\Delta_{\beta}<0$, $\Gamma^\beta_{ij}$ are purely imaginary: $\Gamma_{ij}^\beta|_{\Delta_{\mathrm{\beta}}<0}= i J^{\beta}_{ij}$, with $J^{\beta}_{ij}$ as defined in Eq. \ref{J}. Ultimately, the modified Bessel function $K_0(x)$ is damped by an exponential factor controlled by $\xi_\beta$, that can be tuned dynamically through the detuning $\Delta_{\mathrm{\beta}}$ and made large enough to guarantee that we reach the limit $x=r_ {ij} /\xi_\beta \ll 1$, where $J^\beta_{ij}$ is of strongly long-range character as depicted in Fig. \ref{fig1}(c). In this regime, we engineer then the following general class of XXZ spin hamiltonians:
\begin{equation}
 \label{spin}
H=\sum_{i,j} [J^{z}_{ij} \sigma_i^z\sigma_j^z +J^{xy}_{ij}\ud{\sigma_i}\sigma_j]\,,%-B \sum_{i} \sigma_i^z\,,
\end{equation}
\noindent where $J^{z (xy)}_{ij}$ can be tuned independently by changing the laser intensities, $\Omega_l$ or effective detunings, $\Delta_{\beta}$.

Although conceptually straightforward to place an atomic transition within a band gap for interactions mediated by virtual photons \cite{john90a,shahmoon13a,douglas13a}, this is problematic in practice due to the difficulty of obtaining overlapping band gaps for TE and TM modes in low loss dielectrics in the optical domain. Our configuration relies on the existence of a bandgap for only a single polarization of the structure. It also allows for dynamical tuning of the interaction parameters near the band edge \cite{douglas13a}. 

The strength of atom-atom interactions in the PCW is ultimately determined by $\Gamma_{2d}$ and a logarithmic correction coming from $K_0$ that scales approximately as $\log(d/\xi_\beta)$ and contains the dependence with $\Delta_\beta$.  For the regime $\Delta_{\beta}<0$ (i.e., to obtain purely coherent spin-spin interactions), we find that \cite{SupMat}

\vspace{-2mm}
\begin{align}
 \label{couplingdelta2d2band2}
  \Gamma_{2d}\approx  \Gamma_{\mathrm{a}} \frac{c \sigma}{4\pi A L_m(\omega_c,\mathbf{r}_{\mathrm{a}})}\,,
\end{align}
where $\Gamma_{\mathrm{a}}$ is the free-space radiative decay rate, $\sigma=\frac{3}{2\pi}\eta \lambda^2_{\mathrm{a}}$ is the effective cross section, $\eta$ a correction parameter that depends on the atomic implementation and $L_m (\omega,\mathbf{r}_\mathrm{a})$ the effective mode length that depends on both the atomic position and the electric field density of the GM. In Fig.~\ref{fig1}(c), we plot the scaling of $\Gamma_{2d}$ with the curvature parameter, $A$, using the effective length of the structure in Figs.~\ref{fig3}-\ref{fig4} in the hole center, $L_{m,\mathrm{TE}}\sim 0.3$ $\mu$m, and $\eta=1/2$. The averaged band curvature of our structure in Fig. 3 is $A\approx 1.8\times 10^{12}$ $\mu$m$^2$/s. 
However, to obtain more accurate estimates of $J_{ij}^\beta$ we have performed numerical calculations taking into account the actual anisotropic band structure, as well as details of the atomic implementation, thereby obtaining $J^\beta_{ij}$ of the order $30-40 \Gamma_{\mathrm{a}}$ for detunings such that $\xi_\beta \sim 100 d$~\cite{SupMat}.

Beyond the processes described by Eqs. (4-7), there will be a variety of mechanisms that lead to decoherence for photon-mediated atomic interactions. Absent a full $3$-D band gap for our $2$-D structures, trapped atoms will radiatively decay into free-space and lossy modes within the PCW. To estimate these losses, we have performed FDTD simulations for structures as in Figs.~\ref{fig3}-\ref{fig4}, and find that the total atomic decay rate to all channels except the designated GMs is $\Gamma'\sim 0.4 \Gamma_a$. The PCW itself will have imperfections that can be estimated from the observed quality factors $Q$ for state-of-the-art nanophotonic structures, which recently reported $Q\sim 10^6-10^7$~\cite{srinivasan02a,taguchi11a,sekoguchi14a}. This finite $Q$ translates into a finite photon lifetime, $\kappa=\omega_c/Q$, which together with $\Gamma'$, leads to an effective rate of decoherence $\kappa_{\mathrm{eff}}=\Gamma'+\kappa J_{ij}/\Delta_\beta$ \cite{douglas13a}. Intuitively, the number of spin-exchange 
cycles in the presence of decoherence can be characterized by $\mathcal{N}=J_{ij}/\kappa_{\mathrm{eff}}$. Using our structure as in Figs.~\ref{fig3}-\ref{fig4} and taking $Q=10^7$,  $\mathcal{N}\sim 35$ is obtained for a detuning $\Delta_{\beta}\sim 10$~GHz, which yields $J_{ij}\sim 16 \Gamma_{\mathrm{a}}$ and $\xi_{\beta}\sim 16 d$~\cite{SupMat}. Further improvements in the material quality and alternate lattice geometries could provide flatter bands (i.e., reduced $A$) and better $Q$, thereby increasing both $\mathcal{N}$ and the effective interaction strength and length. 

\paragraph{Conclusion.}

We have shown how $2$-D PCWs can be used to trap atoms and realize new kinds of subwavelength optical lattices with higher tunneling rates for simulations of Bose-Hubbard physics. Moreover, the possibility of combining atom trapping with photon-mediated atom-atom interactions via $2$-D guided modes in PCWs enables realizations of spin models with
large, tunable, and \emph{strong} long-range interactions of both dissipative and coherent character. Beyond the particular examples in Figs. 1-4, we have developed designs for a variety of other structures (e.g., dielectric posts instead of holes and triangular instead of square lattices) and materials (e.g., TiO$_2$, SiC, and SiN) with comparable performance to those described here. Our extensive investigations with diverse structures and materials support the applicability of our projections in Fig.~\ref{fig1}(b-d) to a wide class of problems as in Eqs. 1, 2.

\textbf{Acknowledgements} -
 We gratefully acknowledge discussions with O. Painter. The work of AGT and JIC was funded by the EU integrated project SIQS. AGT also acknowledges support from Alexander Von Humboldt Foundation. JIC acknowledges support as a Moore Distinguished Scholar. DEC acknowledges support from Fundacio Privada Cellex Barcelona. HJK and CLH acknowledge funding by the IQIM, an NSF Physics Frontier Center with support of the Moore Foundation, by the AFOSR QuMPASS MURI, by the DoD NSSEFF program, and by NSF PHY1205729. HJK acknowledges support as an MPQ Distinguised Scholar.

\vspace{-4mm}

 %\bibliography{Sci,books}

%%%%%%%%%% Merge with supplemental materials %%%%%%%%%%

\widetext
\begin{center}
\textbf{\large Supplemental Material: Subwavelength vacuum lattices and atom-atom interactions in photonic crystals}
\end{center}
%%%%%%%%%% Merge with supplemental materials %%%%%%%%%%
%%%%%%%%%% Prefix a "S" to all equations, figures, tables and reset the counter %%%%%%%%%%
\setcounter{equation}{0}
\setcounter{figure}{0}
\makeatletter

%SM for figures
\renewcommand{\thefigure}{SM\arabic{figure}}
\renewcommand{\thesection}{SM\arabic{section}}  
\renewcommand{\theequation}{SM\arabic{equation}}  
%\renewcommand{\bibnumfmt}[1]{[SM#1]}
% \renewcommand{\citenumfont}[1]{SM#1}

% \pacs{42.50.Nn, 73.20.Mf, 71.36.+c}

\section{Planar slab waveguide.}

To gain insight for the design of optical lattices with dielectric structures we consider the simplest structure that supports the existence of `Guided Modes' (GMs) for both polarizations of light: a dielectric slab of width $W$ and refractive index $n(=\sqrt{\varepsilon})$. For illustration we focus on the symmetric transverse electric (TE-like) GM, whose electric profile inside and outside the dielectric is given by;
\begin{align}
 \label{eqTE}
 & \mathbf{E}_{\mathrm{GM},\mathrm{in}}(\mathbf{r})=E_{\mathrm{in}} \cos(k_z z) e^{i k_\| x}\mathbf{y} \,, \nonumber \\
 & \mathbf{E}_{\mathrm{GM},\mathrm{out}}(\mathbf{r})=E_{\mathrm{out}} e^{-\beta z} e^{i k_\| x} \mathbf{y}\,, 
\end{align}
\noindent where $\beta$ and $k_z$ are related by the following trascendental equations \cite{liao91a}:
\begin{align}
 \label{eqTEmodes}
 & \beta = k_z\tan( k_z W/2)\,,\nonumber  \\
 & k_z^2+\beta^2=k_0^2(n^2-1)\, ,
\end{align}
\noindent where $k_0=\omega/c=2\pi/\lambda_0$. The in-plane momentum, $k_\|$, determines the effective wavelength of the GM and can be obtained from the solution ($\beta,k_z$) of the previous equations as $k_\|^2=k_0^2 n^2-k_z^2$. Contrary to conventional optical lattices $k_\|$ depends not only on its wavelength ($\lambda_0$) but also on the slab properties $(n,W)$. By sending counter-propagating GMs in both $x,y$ directions a standing-wave is generated as for conventional optical lattices with intensity periodicity $d_{\mathrm{slab}}=\pi/k_\|$. From Eqs.~\ref{eqTE}-\ref{eqTEmodes}, it can be shown that the upper bound of $k_\|$ yields to a minimum periodicity of: $d_{\mathrm{min},\mathrm{slab}}= \frac{\lambda_0}{2 n}$. This distance is a factor $n$ smaller than the one that can be achieved by interfering lasers in free-space. The electromagnetic profile of this standing wave has associated an optical potential with a periodicity in the $x,y$ plane,  $V_{\mathrm{GM}}(\mathbf{r}+\
\mathbf{R})=V_{\mathrm{GM}}(\mathbf{r})$, where $\mathbf{R}$ is any vector of the Bravais lattice of the potential. However, in the $z$ direction the potential decays exponentially, $e^{-2\beta z}$, due to the evanescent character of the GMs outside the dielectric.

To stabilize the trap in the vertical $z$ direction normal to the slab, Side Illumination (SI) on both sides of the structure can be used. By sending two lasers with the same frequency but whose amplitudes are related by a phase factor $e^{i\phi}$, an interference pattern is generated, whose minima position can be controlled by chirping the phase $\phi$ between them. It is convenient to work in the condition where the slab is transparent to incident light, i.e., when $\lambda_{\mathrm{SI}}=2 n W$. The far off-resonance trapping (FORT) potential associated to SI can be approximated by: 
\begin{equation}
 V_{\mathrm{SI}}(\mathbf{r})=V_{\mathrm{SI}}(z)\approx-\frac{\Omega_{\mathrm{SI}}^2}{\delta_{\mathrm{SI}}}\sin^2\Big[k_{\mathrm{SI}} \big(z-z_t(\phi)\big)\Big] \,,
\end{equation}
where $\Omega_{\mathrm{SI}}=\vec{\mu}\cdot \mathbf{E}_{\mathrm{SI}} /\hbar$ is the Rabi frequency, $\vec{\mu}$ the atomic dipole moment, $ \mathbf{E}_{\mathrm{SI}}$ the electric field associated to SI, and $\delta_{\mathrm{SI}}$ the detuning between the SI and atomic frequencies. The position of the trapping minimum, $z_t$, is controlled through the phase, $\phi$, between up/down lasers.  

The trapping minimum, $z_t$, must be close to the surface to interact with $V_{\mathrm{GM}}$, therefore, Casimir-Polder (CP) potentials must be taken into account \cite{buhmann07a}. In order to estimate their effect over the optically induced trap and under which conditions we can neglect them, we consider the CP potential above a semi-infinite dielectric interface, given by \cite{buhmann07a}:
\begin{align}
 \label{vwdpot1}
V_{\mathrm{CP},\mathrm{plane}}(z)&=-\frac{1}{16}\frac{n^2-1}{
n^2+1}\frac{\Gamma_{\mathrm{a}}}{(k_{\mathrm{a}}z)^3}\,,
\end{align}
where $k_\mathrm{a}$ and $\Gamma_{\mathrm{a}}$ are the atomic momemtum and vacuum decay rate. The stability condition for the trap in the $z$-direction implies $V''_{\mathrm{SI}}(z_t)+V''_{\mathrm{CP},\mathrm{plane}}(z_t)>0$, which can always be satisfied by tuning the intensity of the laser such that

\begin{equation}
 \label{intlas}
 \Omega_{\mathrm{SI}}^2>\frac{3}{2}\frac{\delta_{\mathrm{SI}}V_{\mathrm{CP},\mathrm{plane}}(z_t) }{(k_{\mathrm{SI}} z_t)^2} =\frac{3}{32}\frac{n^2-1}{
n^2+1}\frac{\delta_{\mathrm{SI}}\Gamma_{\mathrm{a}}}{(k_{\mathrm{a}} z_t)^3(k_{\mathrm{SI}} z_t)^2}\,.
\end{equation}

Thus, combining $V_{\mathrm{SI}}(z)+V_{\mathrm{GM}}(\mathbf{r})$, a predominantly FORT optical potential is generated along the $z$ direction with subwavelength lattice constant, $d_{\mathrm{slab}}\gtrsim \frac{\lambda_0}{2 n}$ in the $x,y$ plane. We recognize that the estimate of intensity of Eq. \ref{intlas} using Eq. \ref{vwdpot1} for the CP potential is only a first approximation to the problem, and a full numerical calculation of CP potential has to be considered for more accurate estimations.

\begin{figure*}[!tb]
 \centering
 \includegraphics[width=0.99\linewidth]{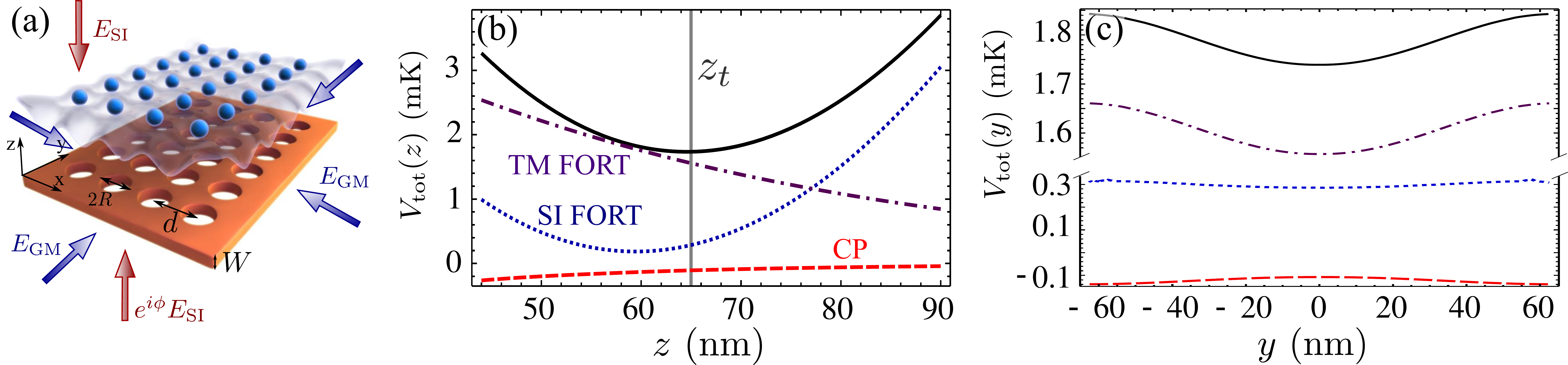}
 \caption{(a) General scheme of the system: dielectric slab of thickness $W$ and refractive index $n$ with periodic lattice of holes of radius $R$ and periodicity $d$. Panels (b-c): Proof of principle example for GaP slab with $d=125$ nm, $R=0.4d$ and $W=2d$. We use an incoherent superposition of two orthogonal TM GM modes for horizontal trapping in the $x,y$ plane with $\lambda_{\mathrm{TM}}=770$ nm and  $\Omega_{\mathrm{GM}}/2\pi= 50$ GHz. For vertical trapping along $z$ we use SI FORT with $\lambda_{\mathrm{SI}}=760$ nm and $\Omega_{\mathrm{SI}}/2\pi= 50$ GHz. (b) [and (c)] Vertical [and horizontal] cut of the total potential (solid black) and its different contributions: CP (dashed red), SI (dotted blue) and TM (dashed-dotted purple) at $x=y=0$ [and $x=0$, $z=z_t=65$ nm], respectively.}
 \label{fig1SM}
 \end{figure*}
 
\section{Photonic Crystal Waveguides (PCWs) for subwavelength trapping.}

Let us now consider the situation where there is a periodic modulation of the refractive index of the structure, $n(\mathbf{r})$, e.g., by assuming that the dielectric slab has a periodic square lattice of circular holes, with radius $R$ and periodicity $d$ [see Fig.~\ref{fig1SM}(a)]-- the so-called Photonic Crystals Waveguide (PCW) slab. The presence of the holes decreases intensity requirements, as the CP potential above the holes is corrected with respect to $V_{\mathrm{plane},\mathrm{CP}}(z)$ \cite{eberlein11a} in the electrostatic regime by a factor,  $f(z,R)$:
 \begin{align}
  \label{vwdpot2}
 V_{\mathrm{hole},\mathrm{CP}}(\mathbf{r})&=V_{\mathrm{plane},\mathrm{CP}}(z)\times f(z,R)=V_{\mathrm{plane},\mathrm{CP}}(z)\times \Big[\frac{1}{2}+\frac{1}{\pi}\arctan\big(\frac{
 z^2-R^2}{2 R z}\big) + \frac{2 z R (z^2-R^2)}{\pi(R^2+z^2)^2}\Big]\,,
 \end{align}
which is always smaller than $1$ above the hole. PCW slabs also support GMs whose properties depend on geometrical/material parameters. The wavelength of the GM is $\lambda_{\mathrm{GM}}=\lambda_0/n'$,  where $n'$ is the effective refractive index of the medium that takes into account geometrical effects. Typically $n' \lesssim n$, thus, the periodic lattice generated by two counter-propagating GMs [see Fig.~\ref{fig1SM}(a)] show periodicities lower bounded approximately by: $d_{\mathrm{min},\mathrm{GM}}\approx \frac{\lambda_0}{2 n}$. SI is used to stabilize the trap in the $z$-direction. At large enough distances $z_t$, the periodic modulation of the CP and SI potential in the $x,y$ plane can be neglected.

In Fig.~\ref{fig1SM}, we show a proof of principle example for a GaP slab with $d=125$ nm, $R=0.4d$ and $W=2d$ which corresponds to the line (i) in Fig. 1(b) of the main manuscript. Curves for the various trapping potentials are calculated numerically without simplifying assumptions as in Eq. (\ref{vwdpot2}). We use an incoherent superposition of two orthogonal TM GM modes for a horizontal trap with $\lambda_{\mathrm{TM}}=770$ nm and  $\Omega_{\mathrm{GM}}/2\pi= 50$ GHz. For vertical trapping we use SI FORT with $\lambda_{\mathrm{SI}}=760$ nm and $\Omega_{\mathrm{SI}}/2\pi=50$ GHz.  Line cuts of the total trap potential $V_{\mathrm{tot}}$ (solid black) are given in Fig.~\ref{fig1SM}(b,c), with the contributions from the TM FORT (dot-dashed purple), SI FORT (dotted blue) and CP (dashed red) potentials shown. The trap depth $V_d$ and frequencies $\omega_{\mathrm{t}}$ for the particular example in Fig.~\ref{fig1SM} are $\{V_{d,\mathrm{xy}},V_{d,\mathrm{z}}\}/2\pi \approx \{2.1,624\}$~MHz and $\{\omega_{t,\mathrm{xy}},
\omega_{t,\mathrm{z}}\}/2\pi\
\approx \{0.52,18.7\}$~MHz.

 \subsection{Contrast loss of GM for deeply subwavelength scales.}

The scheme of the two previous sections use GMs for $x,y$ confinement and SI to trap in the vertical direction. In principle, it is possible to extend this method for deeper subwavelength scales, $d\ll \lambda_0/(2 n)$; however, several complications arise in this limit. For example, as the lattice constant $d$ decreases, the GM with energies around the atomic transition are closer to the light line. Consequently, the contrast of the GM intensity in the $x,y$ plane for a given trap distance $z_t$, that can be characterized through the function $C(z_t)=(\mathrm{max}\{|E_{\mathrm{GM}}(x,y,z_t)|^2\}-\mathrm{min}\{|E_{\mathrm{GM}}(x,y,z_t)|^2\})/(\mathrm{max}\{|E_{\mathrm{GM}}(x,y,z_t)|^2\}+\mathrm{min}\{|E_{\mathrm{GM}}(x,y,z_t)|^2\})$, also decreases. This contrast loss results in the necessity of larger laser intensities to guarantee the trapping condition, $2\omega_t\lesssim V_d$. The FORT trap depth and frequency scale as $V_d =C(z)|\Omega_{\mathrm{GM}}|^2/|\delta|$ and $\omega_t =\sqrt{h  V_d /(2 m d^2)}$, 
respectively, where $\Omega_{\mathrm{GM}}$ is the Rabi frequency associated to the GMs. Theoretically, the trapping condition can always be satisfied by compensating the contrast loss with the Rabi frequency strength. Concretely, if we denote $\Omega_{\mathrm{conf}}$ to be the value that guarantees the equality [$2\omega_t= V_d$], $\Omega_{\mathrm{conf}}$ is found to be
  \begin{align}
   \label{key}
   \Omega_{\mathrm{conf}}\simeq\sqrt{\frac{2 h c |\Delta\lambda^{-1}|}{m d^2 C}}\,,
  \end{align}
  where $\Delta\lambda^{-1}=1/\lambda_{\mathrm{a}}-1/\lambda_{\mathrm{GM}}$.

\begin{figure*}[!hb]
 \centering
 \includegraphics[width=0.7\linewidth]{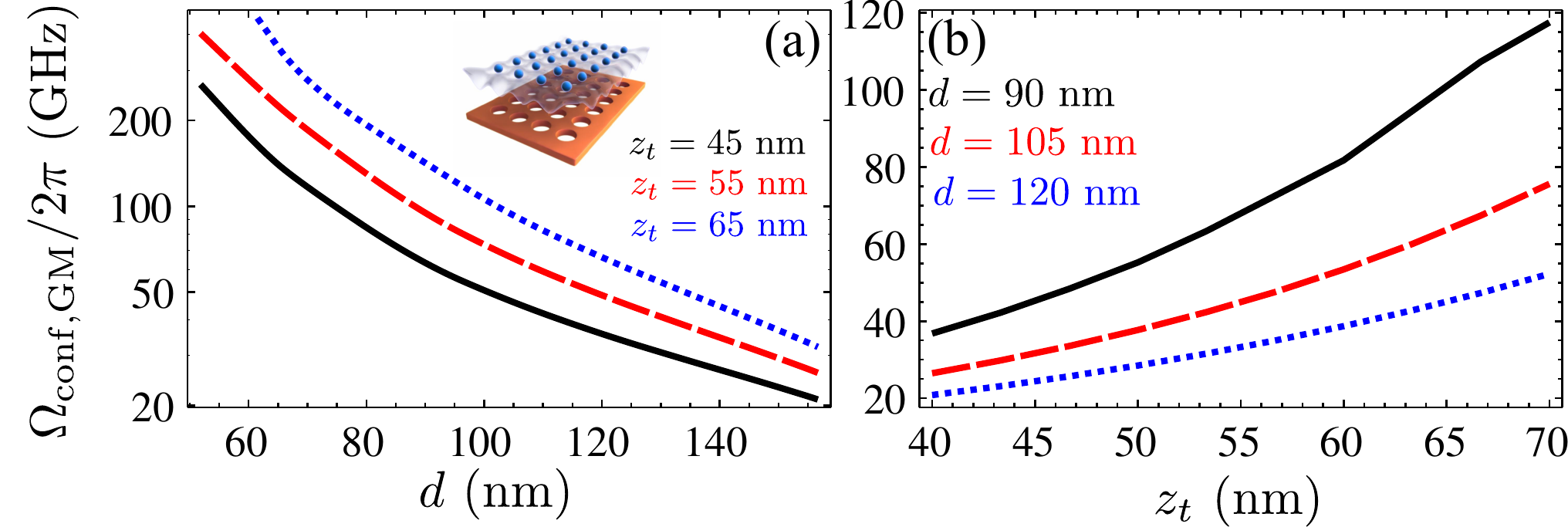}
 \caption{(a) $\Omega_{\mathrm{conf}}$ for a GaP PCW with $W=2d$ and $R=0.4d$, as defined in Eq. \ref{key}, as a function of lattice parameter $d$ for different vertical distances $z_t$ from the surface, as depicted in the legend. (b)  $\Omega_{\mathrm{conf}}$ for a GaP PCW with $W=2d$ and $R=0.4d$, as defined in Eq. \ref{key}, as a function of vertical distance from the surface, $z_t$, for different lattice parameters $d$ as depicted in the legend. }
 \label{fig3sup}
 \end{figure*}

 In Fig.~\ref{fig3sup}, we show an example of these scalings for a GaP slab with $W=2d$, $R=0.4 d$, considering a fixed GM wavelength of $\lambda_{\mathrm{GM}}=760$ nm. From the study of the horizontal confinement provided by the GMs, we see that: i) The smaller the lattice parameter $d$, the smaller is the contrast of the GM modes, implying a considerable increase of the intensities required to reach the trapping condition in the $x,y$ plane (see Fig.~\ref{fig3sup}(a)); ii) For a fixed lattice constant parameter $d$, it is possible to decrease the intensity requirements by choosing a smaller vertical trapping position, $z_t$, as shown in panels (a-b) of Fig.~\ref{fig3sup}. However, this strategy requires increasing SI intensities to compensate the CP potential (which is not included in Eq. \ref{key} nor in Fig. \ref{fig3sup}).  Summing up, obtaining deep subwavelength scales using GM (and SI) for horizontal (vertical) trapping can be done at the expense of significant increases in the laser intensities with 
the 
concomitant increase in the scattering and heating rates.

\section{Scalings in ``vacuum lattices''.} 

\begin{figure}[!tb]
 \centering
 \includegraphics[width=0.99\linewidth]{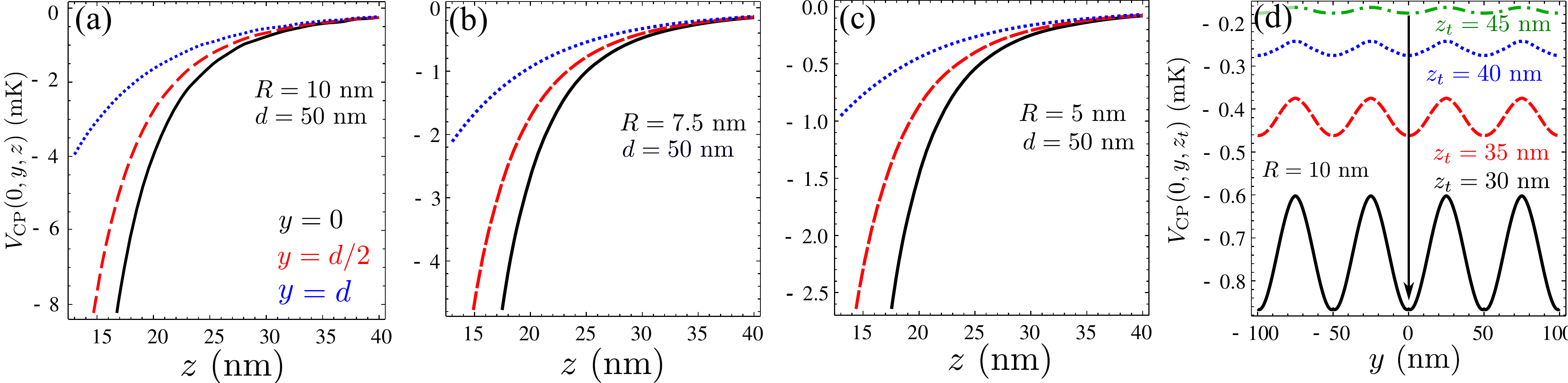}
 \caption{Panel (a,b,c): Vertical cut of the CP potential $V_{\mathrm{CP}}(0,y,z)$ for three different horizontal positions in the unit cell: $y=0$ (solid black), $y=d/4$ (dashed red) and $y=d/2$ (dotted blue) for a structure with $d=50$~nm and $R=10$~nm (a), $R=7.5$~nm (b) and $R=5$~nm (c). Panel (d): Horizontal cut of the CP potential for a periodic lattice of GaP posts of $R=10$~nm and $d=50$ nm for different vertical distances, $z_t$, ranging from $45$ to $30$~nm. }
 \label{fig4sup}
  \vspace{-4mm}
 \end{figure}

In the main manuscript we show how to engineer a new class of optical lattices taking advantage of the CP modulation in the XY-direction. In these ``vacuum lattices'' the XY confinement is mainly provided by CP, whereas the vertical confinement comes from SI.  In Fig.~\ref{fig4sup} (a-c) we show how the CP potential varies in the vertical direction for different places of the unit cell and different post radii. Here, the unit cell is centered at the position of the vertical axis of the dielectric post. $x=y=0$. As expected the CP potential is larger in absolute value in the regions close to the dielectric, $y=0$, than along the border of the unit cell $y=d/2$. Moreover, the case with the biggest post radius, $R=10$ nm, also shows deeper potentials in the XY plane. The dependence of the CP for the XY confinement as a function of the trapping distance $z_t$ is summarized in panel (d) of Fig.~\ref{fig4sup}.  The advantage of the CP potential compared to $V_{\mathrm{GM}}$ is that it does not suffer from the 
contrast loss for smaller 
distances $\{d,z_t \}
\ll \lambda_0$.

\section{Atom-atom interactions $\Gamma_{ij}$ mediated by Guided modes.}

When atoms are trapped close to the PCW structures, they interact with the GMs in the structure. This interaction can be described by the following hamiltonian~\cite{john90a,john94a,john95a}:
\begin{equation}
 \label{hint1}
 H_{int}=\sum_{i}\sum_{\mathbf{k}}\big( g_\mathbf{k} e^{i\mathbf{k}\cdot
\mathbf{r}_i}a_{\mathbf{k}}\ud{\sigma_i}+\mathrm{h.c.}\big)\,,
\end{equation}
where we neglected the multi-band effects (as we work with $\Delta_{\mathrm{a}c}=\omega_{\mathrm{a}}-\omega_c$ much smaller than the bandgap width) and focused on a single polarization. The coupling constant can then be written as:
\begin{equation}
\label{coup}
 g_{\mathbf{k}}=\sqrt{\eta\frac{\omega(\mathbf{k})}{2 \varepsilon_0 \hbar L^2}} \vec{\mu}\cdot \mathbf{u}_{\mathbf{k}}\,,
\end{equation}
where $\vec{\mu}$ is the atomic dipole moment, $L^2$ is the quantization area, $\omega(\mathbf{k})$ [$\mathbf{u}_{\mathbf{k}}$] is the energy dispersion [spatial mode dependence] of the field $a_{\mathbf{k}}$, and $\eta$ a factor that takes into account polarization effects for the particular atomic level structure addressed. The hamiltonian of Eq.~\ref{hint1} induces an effective interaction between the atoms. If the coupling between atoms and dielectric modes can be treated under the Born-Markov approximation, it is then possible to obtain a master equation that describes effectively the dynamics of the atoms by tracing out the photonic degrees of freedom. The effective equation is then given by \cite{breuerbook02a}:
\begin{equation}
 \label{efmaster2}
 \frac{d\rho}{dt}=\sum_{i,j} \Gamma_{ij}\big( \sigma_i\rho\ud{\sigma_j}
-\ud{\sigma_j}\sigma_i\rho\big)+\mathrm{h.c} \,,
\end{equation}
\noindent where $\Gamma_{ij}$ is the collective coupling defined as follows:

\begin{align}
 \label{coupling}
 \Gamma_{ij}= L^2\lim_{s\rightarrow 0} \int_{\mathrm{BZ}}\frac{d^2 \mathbf{k}}{(2\pi)^2}\frac{|g_{\mathbf{k}}|^2}{s+i(\omega_{\mathrm{a}}-\omega(\mathbf{k}))}e^{i \mathbf{k}\cdot \mathbf{r}_{ij} }\,,
\end{align}
where $L^2$ is the quantization area and the $\mathbf{k}$-integration is over the entire Brillouin zone.

We can separate in Eq.~\ref{efmaster2} the contributions from the real and imaginary parts of $\Gamma_{ij}$. Denoting: $\Gamma_{ij}=\gamma_{ij}/2+i J_{ij}$, where both $\gamma_{ij}$ and $J_{ij}$ are real numbers, Eq.~\ref{efmaster2} can be rewritten as follows:

\begin{equation}
 \label{efmaster22}
 \frac{d\rho}{dt}=\sum_{i,j} \frac{\gamma_{ij}}{2}\big( 2\sigma_i\rho\ud{\sigma_j}
-\ud{\sigma_j}\sigma_i\rho-\rho\ud{\sigma_j}\sigma_i\big)+i\sum_{ij} J_{ij}[\rho,\ud{\sigma_j}\sigma_i]\,,
\end{equation}
which is separated into the coherent (incoherent) contribution coming from $J_{ij}$ ($\gamma_{ij}$), respectively. 

\begin{figure}[!tb]
 \centering
 \includegraphics[width=0.89\linewidth]{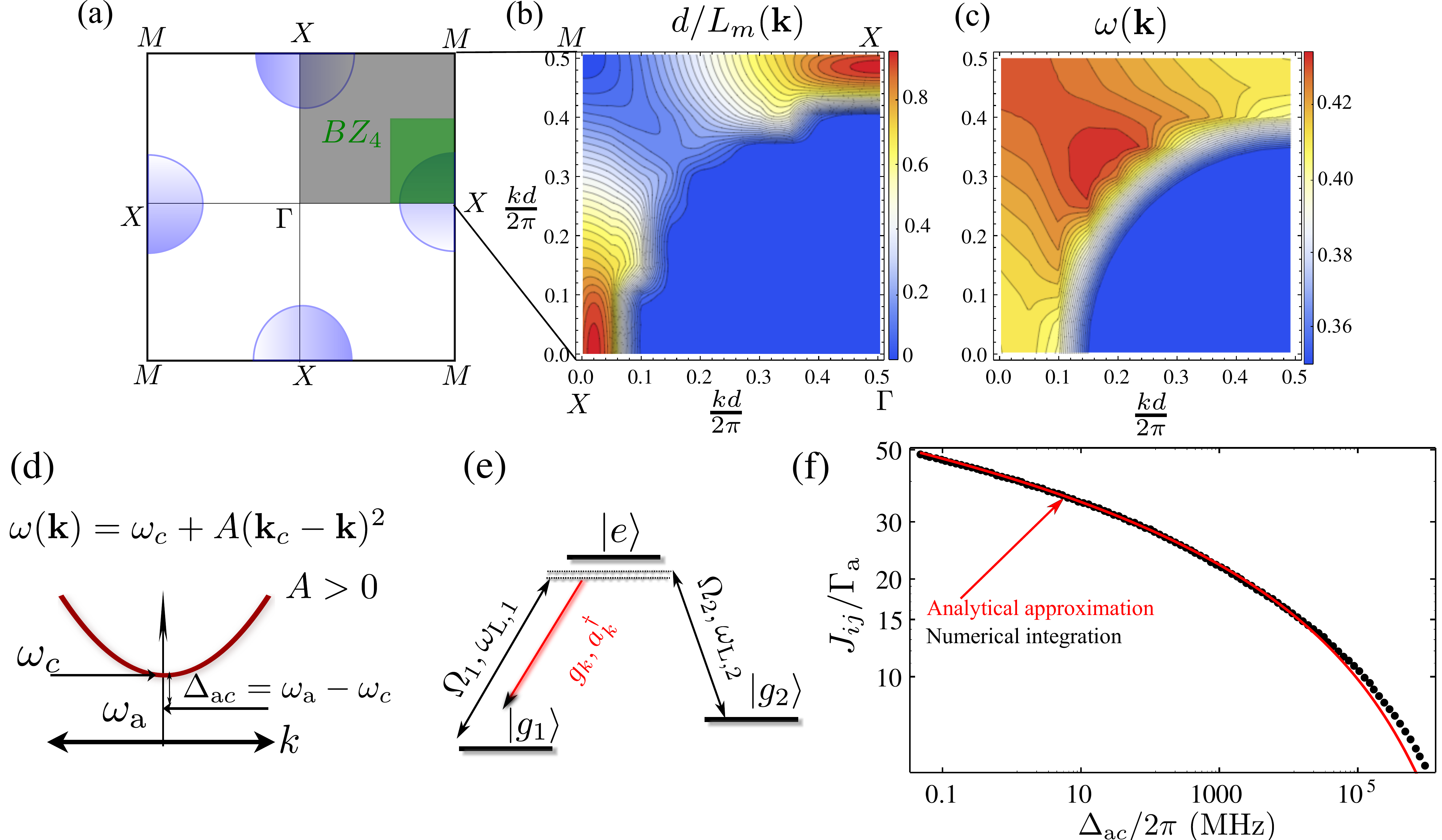}
 \caption{(a) Brillouin zone highlighting in blue the four most relevant regions contributing to $\Gamma_{ij}$. (b) Inverse of effective mode length, $d/L_m(\mathbf{k})$, and (c) energy dispersion for the square hole lattice structure of Figs.~ 3-4 of Ref. \cite{gonzaleztudelaMain}. (d) Parabolic approximation of the band structure close to a band edge. (e) Simplified scheme in order to couple to a single polarization of light. (f) Coherent coupling, $J_{ij}/\Gamma_{\mathrm{a}}$, for $r_{ij}=d$ using the exact numerical integration and the analytical expression obtained using an isotropic approximation and an ``averaged'' $A_{\mathrm{fit}}\sim 1.8\times 10^{12} $ $\mu$m$^2/s$. The ``averaged'' value of $A$ is chosen to fit the numerical integration data (in solid black).}
 \label{fig6sup}
 \end{figure}

For the structure considered in Figs.~3-4 in Ref.~\cite{gonzaleztudelaMain}, the main contributions of the integration over the Brillouin zone in Eq. \ref{coupling} are given by the regions within the four semi-circles depicted in Fig.~\ref{fig6sup}(a) for the four $X$ points of the Brillouin zone. Separating the different contributions, we arrive at:

\begin{align}
 \label{coupling3}
\Gamma_{ij}=\Big[\Gamma_{ij,x} \cos(\mathbf{k}_{c,x}\cdot \mathbf{r}_{ij})+\Gamma_{ij,y} \cos(\mathbf{k}_{c,y}\cdot \mathbf{r}_{ij}) \Big] \,,
\end{align}
where $\Gamma_{ij,x (y)}$:
\begin{align}
 \label{coupling4}
\Gamma_{ij,x (y)}=L^2\lim_{s\rightarrow 0} \int_{\mathrm{BZ}}\frac{d^2 \mathbf{k}}{(2\pi)^2}\frac{|g_{\mathbf{k}_{c,{x,y}}-\mathbf{k}}|^2}{s+i(\omega_{\mathrm{a}}-\omega(\mathbf{k}_{c,x(y)}-\mathbf{k}))}e^{i \mathbf{k}\cdot \mathbf{r}_{ij} } \,.
\end{align}
The atoms are placed at positions $\mathrm{r}_{ij}=(n,m)d$, with $n,m\in \mathbb{Z}$ such that the cosine terms only give phases: $(-1)^n,(-1)^m$. These phases could also be compensated by an appropriate configuration of the in-plane momenta of the driving lasers in our two-photon Raman coupling scheme (see Fig.~\ref{fig6sup}(d)). As we are interested in calculating the magnitude of $\Gamma_{ij}$, for simplicity, we drop these phases and also assume that the coupling is symmetric for the X and Y directions ($\Gamma_{ij,x}=\Gamma_{ij,y}$). Then, the absolute value of coupling $\Gamma_{ij}$ can be estimated by calculating twice the integral of the circumference area around the X point:
\begin{align}
 \label{coupling5}
\Gamma_{ij}=2 L^2\lim_{s\rightarrow 0}  \int_{BZ}\frac{d^2 \mathbf{k}}{(2\pi)^2}\frac{|g_{\mathbf{k}_{c}-\mathbf{k}}|^2}{s+i(\omega_{\mathrm{a}}-\omega(\mathbf{k}_{c}-\mathbf{k}))} e^{i \mathbf{k}\cdot \mathbf{r}_{ij} } \,.
\end{align} 

 \subsection{Using ``effective mass'' for $\omega(\mathbf{k})$ and isotropic approximation.}

Up to this point we have not made assumptions about $g_{\mathbf{k}}$ or $\omega(\mathbf{k})$. In order to obtain analytical expressions that give us some intuition on the magnitude and scaling of $\Gamma_{ij}$, we use the effective mass approximation: $\omega(\mathbf{k})\approx \omega_{c}+A (\mathbf{k}_c-\mathbf{k})^2$ [see Fig.~3 in Ref.~\cite{gonzaleztudelaMain}], and assume that the coupling is isotropic around the X-points. Using these assumptions, we integrate the angular dependence in the integral of Eq.~\ref{coupling5} as follows:
\begin{align}
 \label{coupling9}
\Gamma^{\mathrm{iso}}_{ij}= \lim_{s\rightarrow 0} \frac{L^2 |g_{\mathbf{k}_{c}}|^2}{\pi}\int_0^{k_c}dk  \frac{k J_0(k |r_{ij}| )}{s+i(\Delta_{\mathrm{a}c}-A k^2)}\,,
\end{align}
where we have taken  $|g_{\mathbf{k}_{c}-\mathbf{k}}|^2\rightarrow |g_{\mathbf{k}_{c}}|^2$ out of the integral for a simple model. For $A>0$ there are two different situations to consider, namely,  $\Delta_{\mathrm{a}c}\lessgtr 0$. When the atomic frequency lies in the bandgap, $\Delta_{\mathrm{a}c}<0$, the limit has no singularity and the $\Gamma^{\mathrm{iso}}_{ij}=-i J^{\mathrm{iso}}_{ij}$ is purely imaginary, with:
\begin{align}
 \label{coupling9}
J^{\mathrm{iso}}_{ij}=\frac{L^2 |g_{\mathbf{k}_{c}}|^2}{\pi}\int_0^{k_c}dk  \frac{k J_0(k |r_{ij}| )}{\Delta_{\mathrm{a}c}-A k^2}\,.
\end{align}

Defining $\xi=\sqrt{|A/\Delta_{\mathrm{a}c}|}$ and performing the change of variables $q=k\xi$, we arrive to:

\begin{align}
 \label{coupling99}
J^{\mathrm{iso}}_{ij}= \frac{|g_{\mathbf{k}_{c}}|^2 L^2}{\pi A}\int_0^{k_c\xi}dq q \frac{1}{1+q^2} J_0(q |r_{ij}|/\xi )\simeq \frac{ |g_{\mathbf{k}_{c}}|^2 L^2}{\pi A}K_0(|r_{ij}|/\xi)\equiv\Gamma_{2d} K_0(|r_{ij}|/\xi) \,,
\end{align}
where we have used $k_c\xi\gg 1$.  $K_0(x)$ is a modified Bessel function. Defining the free-space decay rate as $\Gamma_{\mathrm{a}}=\frac{|\mu|^2 k_{\mathrm{a}}^3}{3\pi \varepsilon_0 \hbar}$, we write the coupling $|g_{\mathbf{k}}|^2$ as follows:
 \begin{equation}
\label{couplomega}
 |g_{\mathbf{k}}|^2=\Gamma_{\mathrm{a}}\frac{c \sigma \omega(\mathbf{k})}{8 L^2  L_m(\mathrm{k},\mathbf{r}_{\mathrm{a}}) \omega_{\mathrm{a}}} \,,
\end{equation}
where we have defined $\sigma=\frac{3 \eta}{2 \pi} \lambda^2_{\mathrm{a}}$ as the effective cross-section. Substituting the $|g_{\mathbf{k}_{c}}|^2$ into Eq.~\ref{coupling99}, we arrive to:
\begin{align}
 \label{coupling10}
\Gamma_{2d}= \Gamma_\mathrm{a}\frac{c \sigma }{4\pi A L_m (\mathbf{k}_c,\mathbf{r}_\mathrm{a})}\,,
\end{align}
where we have introduced the effective mode length, $L_{m}(\mathbf{k},\mathbf{r}_\mathrm{a})$ defined as:
 \begin{equation}
  \label{effectivelength}
  L_m (\mathbf{k},\mathbf{r}_\mathrm{a})=\frac{\int d^3\mathbf{r}\epsilon(\mathbf{r}) |\bar{E}_{\mathbf{k},m}(\mathbf{r})|^2}{d^2\epsilon(\mathbf{r}_{\mathrm{a}}) |\bar{E}_{\mathbf{k},m} (\mathbf{r}_{\mathrm{a}})|^2}\,,
 \end{equation}
that takes into account the geometrical distribution of the field density, $|\bar{E}_{\mathbf{k},m} (\mathbf{r}_{\mathrm{a}})|^2$, at the atomic position $\mathbf{r}_{\mathrm{a}}$.

In the opposite limit, i.e., the dissipative regime, when $\Delta_{\mathrm{a}c}>0$ the limit $s\rightarrow 0$ has a singularity, that can be calculated using:
\begin{equation}
\label{limit}
\lim_{s\rightarrow 0} \frac{1}{s+i\Delta_k }=\pi \delta(\Delta_k)-i
\mathcal{P}\frac{1}{\Delta_k}\,.
\end{equation}
Therefore in this case $\Gamma_{ij}$ has both real and imaginary components and is given as follows:
\begin{align}
 \label{couplingdelta}
 \Gamma^{\mathrm{iso}}_{ij}\approx \Gamma_{2d}\frac{\pi}{2} H^{(1)}_0 \big[|r_{ij}|/\xi\big]\,,
\end{align}
where $H^{(1)}_0(x)=J_0(x)+iY_0(x)$ is the Hankel function of the first kind.

\subsection{Collective coherent coupling $J_{ij}$ using exact $\omega(\mathbf{k})$ and $g_{\mathbf{k}}$.}

The effective mass and isotropic approximation are convenient to obtain analytical formulas to understand scaling with parameters such as $A$, $\Delta_{\mathrm{a}c}$ or $r_{ij}$. For the structure of Figs. (3-4) of Ref.~\cite{gonzaleztudelaMain}, both the exact dispersion relation $\omega(\mathbf{k})$, and its effective length at the center of the hole, $L_m(\mathbf{k})$, are anisotropic [see Panels (b-c) of Fig.~\ref{fig6sup}]. For example, the curvature (effective length) along the $X-M$ direction is flatter (steeper) than in the $X-\Gamma$ direction. Therefore, it is not straightforward to determine an appropiate value of $A$ to input into the isotropic formula of Eq.~\ref{coupling10}. Moreover, we should also take into account the details of the implementation of the simplified $\Lambda$-scheme depicted in Fig.~\ref{fig6sup}(e). 

To estimate up to which point the isotropic formula is giving the right scalings and order of magnitude for $\Gamma_{ij}$, we numerically integrate Eq.~\ref{coupling5} with the exact $\omega(\mathbf{k})$ and $g_{\mathbf{k}}$ extracted from band structure calculation with the parameter of Figs. (3-4) of Ref.~\cite{gonzaleztudelaMain}. We focus on the situation when the atomic transition lies in the bandgap, $\Gamma_{ij}=i J_{ij}$. To optimize the number of points of numerical integration, we integrate over one quadrant around the X point (that we denote by $BZ_4$ as schematically depicted in Panel (a)) and multiply by 4 to consider the contribution of the whole area; that is, we take:

\begin{align}
 \label{coupling15}
J_{ij}=8 L^2\lim_{s\rightarrow 0}  \int_{BZ_4}\frac{d^2 \mathbf{k}}{(2\pi)^2}\frac{|g_{\mathbf{k}_{c}-\mathbf{k}}|^2}{s+i(\omega_{\mathrm{a}}-\omega(\mathbf{k}_{c}-\mathbf{k}))} \cos(\mathbf{k}\cdot \mathbf{r}_{ij})\,.
\end{align} 

By choosing an atomic position such that $|\mathbf{r}_{ij}|=d$, we obtain $J_{ij}/\Gamma_a \sim 30$  for detunings $\Delta_{\mathrm{a}c}/2\pi\sim 30$ MHz. The scaling with $\Delta_{\mathrm{a}c}$ is logarithmic as predicted by $K_0(r_{ij}/\xi)$ using the isotropic approximation [see Fig.~\ref{fig6sup}(g)]. The best fit of $A$ (such that the isotropic formula and the numerical integration match) is $A_{\mathrm{fit}}\sim 1.8\times 10^{12} $ $\mu$m$^2$/s (using $\eta=1/2$), which is closer to the curvature along the $X-M$ direction, $A_{X-M}\sim 1.5\times 10^{12} $ $\mu$m$^2$/s, than along the $X-\Gamma$ direction, $A_{X-\Gamma}\sim 1.1\times 10^{13} $ $\mu$m$^2$/s. This is because the flatter direction is also more weighted by $1/L_m$ than the one in $X-\Gamma$.

\subsection{Ratio between coherent and incoherent processes.}

In the previous Section, we estimated the rate of the coherent processes in our system ($J_{ij}$). However, there will be several factors that limit the coherence of our system. The absence of a complete $3$-D bandgap or the presence of other polarization modes yields a decay into other radiative channels, that we estimated using FDTD calculations (not shown) to be $\Gamma'\approx 0.4 \Gamma_{\mathrm{a}}$ for our structure in Figs. 3 and 4 in Ref. \cite{gonzaleztudelaMain}.

Moreover, the fabrication process of these structures results in imperfections of the photonic crystal that yields scattering of the guided photons to the non-guided ones at a rate $\kappa$, that is characterized by the so-called quality factor $Q=\omega_c/\kappa$. For square lattices of  $15 \times 15$ holes $Q\sim 10^3-10^4$ has been reported~\cite{kwon03a,cho05a}. Further improvements in the material and fabrication quality yield higher $Q\sim 10^6-10^7$ as recently reported~\cite{srinivasan02a,taguchi11a,sekoguchi14a}.

\begin{figure}[!tb]
 \centering
 \includegraphics[width=0.99\linewidth]{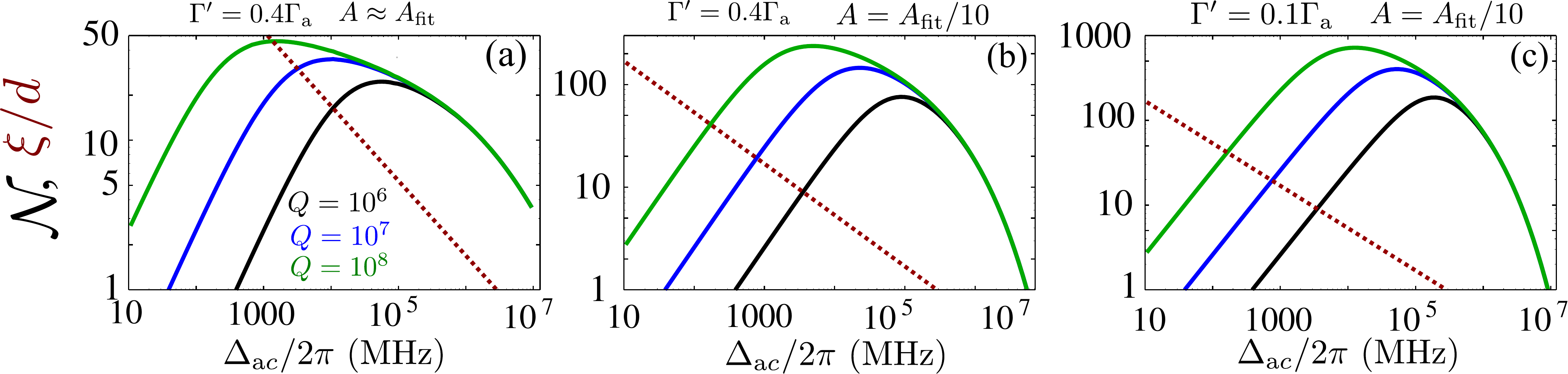}
 \caption{(a) Number of coherent cycles $\mathcal{N}$ as a function of detuning $\Delta_{\mathrm{ac}}$ for the structure of the main manuscript, $\Gamma'=0.4\Gamma_{\mathrm{a}}$ and $Q=10^6$ (black), $Q=10^7$ (blue) and $Q=10^8$ (green). Dashed red line correspond to the normalized effective interaction length $\xi/d$ (b) Same as (a), but using $J_{ij}^\mathrm{iso}$ with a reduced curvature, $A=A_{\mathrm{fit}}/10$, with respect to the structure of the main manuscript. (c) Same as (b), but with reduced $\Gamma'=0.1\Gamma_{\mathrm{a}}$.}
 \label{fig7sup}
 \end{figure}

It can be shown~\cite{douglas13a} that the error rate introduced by both $\kappa$ and $\Gamma'$ within the exchange time of an excitation between two spins ($1/J_{ij}$) through the off-resonant atom-induced cavity is given by $\kappa_{\mathrm{eff}}=\kappa J_{ij}/\Delta_{\mathrm{ac}}+\Gamma'$. In order to quantify the ratio between coherent and incoherent processes, we define:

\begin{equation}
 \label{coop}
 \mathcal{N}=\frac{J_{ij}}{\kappa_{\mathrm{eff}}}=\frac{J_{ij}}{\kappa J_{ij}/\Delta_{\mathrm{ac}}+\Gamma'}\,,
\end{equation}
which intuitively represents the number of cycles of the coherent exchange before a non-desired transition occurs (e.g., a decay event into a continuum of unguided modes). In Fig.~\ref{fig7sup}, we plot $\mathcal{N}$ for state of the art qualify factors $Q= 10^6,10^7$, as well as a projection to $Q=10^8$. We also consider improvements in design and fabrication of the PCW which might lead to reduced band curvature $A$ and radiation loss $\Gamma'$. We find that there is an optimal detuning that maximizes $\mathcal{N}$, i.e., when $\Delta_{\mathrm{ac}}\simeq\kappa J_{ij}/\Gamma'$. In Fig.~\ref{fig7sup}(a) we plot $\mathcal{N}$ for the structure considered in the main manuscript, using $\Gamma'=0.4\Gamma_{\mathrm{a}}$, and $Q=10^6$ (black), $10^7$ (blue) and $10^8$ (green). Using $Q=10^7$, the maximum $\mathcal{N}\sim 35$ is obtained for a detuning $\Delta_{\beta}\sim 10$~GHz, which yields $J_{ij}\sim 16 \Gamma_{\mathrm{a}}$ and $\xi_{\beta}\sim 16 d$.

In Fig.~\ref{fig7sup}(b), we explore the effect of a reduced curvature $A=A_{\mathrm{fit}}/10$ using  $J_{ij}^\mathrm{iso}$ defined in the previous section, by keeping $\Gamma'=0.4\Gamma_{\mathrm{a}}$, obtaining approximately a two-fold enhancement of $\mathcal{N}$. Finally, we explore the effect of reducing $\Gamma'$ to $0.1\Gamma_{\mathrm{a}}$ in panel (c), obtaining approximately a five-fold enhancement of $\mathcal{N}$, being able to reach $\mathcal{N}\sim O(10^3)$ for $Q\sim 10^9$.

\subsection{Engineering with a general $\Lambda$ scheme. }

Here we derive an effective hamiltonian describing the interaction between an effective spin level formed by two metastable atomic states and the guided modes of the structure by using a $\Lambda$-level  as shown in Fig.~\ref{fig6sup}(e). This implementation has several advantages: i) it allows to engineer more complex hamiltonians, e.g., XXZ spin hamiltonians; ii) the Raman process effectively narrows the natural linewidth of the excited state keeping the cooperativity of the process constant \cite{douglas13a}; iii) it requires only a single polarization bandgap, relaxing the requirements in lattice geometry and index contrast. By going to a rotating frame of the driving fields, the complete hamiltonian, with the notation of Fig.~\ref{fig6sup}(e), is given by:

\begin{align}
 \label{hamlaserLi}
H=H_0+H_L+H_I=(\omega_{g1}+\omega_{L,1}) \ket{g_1}\bra{g_1} +(\omega_{g2}+\omega_{L,2})  \ket{g_2}\bra{g_2}+\omega_{e} \ket{e}\bra{e}+\sum_k \omega_k \ud{a_k} a_k +\nonumber \\  \frac{\Omega_1}{2}\big(\ket{g_1}\bra{e} +\mathrm{h.c.}\big)+\frac{\Omega_2}{2}\big(\ket{g_2}\bra{e} +\mathrm{h.c.}\big)+\sum_k g_k \big(\ud{a_k} \ket{g_1}\bra{e} e^{-i \omega_{L,1} t}+\mathrm{h.c.}\big)\,,
\end{align}
which has three different contributions: the free energy, $H_0$, the coherent driving of the lasers, $H_L$, and the interaction with the modes of the structure, $H_I$. We define the following detunings: $\Delta_i=\omega_e-(\omega_{g_i}+\omega_{L,i})$ and apply the following transformation: $ U=e^S=e^{\frac{\Omega_1}{2 \Delta_1 }(\ket{e}\bra{g_1}-\ket{g_1}\bra{e})+\frac{\Omega_2}{2 \Delta_2 }(\ket{e}\bra{g_2}-\ket{g_2}\bra{e})}$. When $|\Delta_i|\gg \Omega_{i}$, the excited state is only virtually populated and the result of the transformation, $H\rightarrow e^S H e^{-S}$, yields in lower order of $\Omega_i/\Delta_i$ to a hamiltonian with different terms: $H=H_0+H_1+H_2+H_3+H_4$. The first one is the free energy hamiltonian, $H_0$, as in the first line of Eq.~\ref{hamlaserLi}. The second and third terms, $H_1, H_2$, are the Stark-shifts and coherent driving of the ground-state levels induced by the off-resonant driving:
\begin{align}
 H_1=-\frac{\Omega_1^2}{4\Delta_1} \ket{g_1}\bra{g_1} -\frac{\Omega_2^2}{4\Delta_2}  \ket{g_2}\bra{g_2}\,, \\
 H_2=-\frac{\Omega_1 \Omega_2}{4}\Big(\frac{1}{\Delta_1}+\frac{1}{\Delta_2}\Big) \big(\ket{g_1}\bra{g_2} +\ket{g_2}\bra{g_1}\big)\,.
\end{align}

$H_1$ can be considered as a renormalization of the free energies where $H_2$ describes Raman processes. Finally, defining an effective qubit with $\{\ket{g_1},\ket{g_2}\}$, such that $\sigma=\ket{g_1}\bra{g_2}$, $\ud{\sigma}=\ket{g_2}\bra{g_1}$ and $\sigma_z=\ket{g_1}\bra{g_1}+1/2$, the transformation of the $H_I$ yields to \cite{footnote1}:

\begin{align}
 \label{hamlaserLi4}
 H_3=-\sum_k g_k \frac{\Omega_2 }{2\Delta_2} \big(\ud{a_k} \sigma e^{i(\omega_k-\omega_{g_2}+\omega_{g_1} -\omega_{L,2} )t}  +\mathrm{h.c.}\big)\,, \\
 H_4=-\sum_k g_k \frac{\Omega_1 }{2\Delta_1} \big(\ud{a_k}  \sigma^z e^{i (\omega_k-\omega_{L,1} ) t} +\mathrm{h.c.}\big)\,.
\end{align}

\noindent which are the most relevant terms for our discussion, as they contain the interaction with the GMs of the structure ($a_k$). By extending these effective light-matter hamiltonians to many atoms, and adiabatically eliminating the GMs in the conditions where the atomic transitions lie in the bandgap leads to the following XXZ spin hamiltonian $H_{\mathrm{spin}}$ from Eqs. SM29-SM30:
\begin{equation}
 \label{spin}
H_{\mathrm{spin}}=\sum_{i,j} [J^{z}_{ij} \sigma_i^z\sigma_j^z +J^{xy}_{ij}\ud{\sigma_i}\sigma_j]\,.%-B \sum_{i} \sigma_i^z\,,
\end{equation}
where $J^{\beta}_{ij}=h_{\beta} \Gamma_{2d} K_0(r_{ij}/\xi_\beta)$, where we have used the effective mass and isotropic approximation (see previous Section). The factor $h_\beta=\big(\Omega_l/(2\Delta_{\beta})\big)^2$ with $l=1\,(2)$ for $\beta=z\, (xy)$ can be controlled independently for each $\beta$ through the laser intensities, $\Omega_l$, and detunings, $\Delta_l$. The effective length scale, $\xi_\beta=\sqrt{A/\Delta_{\beta}}$, depends on both the curvature of the band, $A$, and effective detuning,  $\{\Delta_{xy}, \Delta_z\}  \lessgtr 0$, where $\Delta_{xy}=\omega_{g,2}-\omega_{g,1}+\omega_{L,2}-\omega_c$ and $\Delta_z=\omega_{L,1}-\omega_c $. Therefore, both the length scale, through $\xi_\beta$, and the strength of the interactions, through $h_\beta$, can be tuned independently for each $J^{\beta}_{ij}$-component.

\section{GaP: Material Properties and Fabrication. }

Our analysis in Ref. \cite{gonzaleztudelaMain} considers Gallium Phosphide (GaP) for the various examples presented. We have made this choice in order to present the most favorable case for sub-wavelength optical traps in nanophotonic dielectric waveguides. GaP is a high index ($n \simeq 3.25$), low loss III-IV semiconductor with an indirect band gap at $550$~nm. For other dielectrics transparent in the frequency range of electronic transitions for alkali atoms (e.g., SiN, SiC, TiO$_2$), the lattice constant $d_{\mathrm{min}}=\lambda_0/2n$ would be larger than for GaP (e.g., $n_{SiN}=2.0$, $n_{SiC}=2.6$, and $n_{TiO_2}=\{2.5.2.8\}$ for ordinary and extraordinary polarizations, respectively), even if these materials may have other more favorable properties (e.g., lower absorption) and more advanced processing capabilities. In fact, our schemes for vacuum trapping (Fig. 2) and photon-mediated atomic interactions (Figs. 3, 4) do not rely on such a high-index material as GaP. 

That said, here we gather relevant information related to material properties and state-of-art fabrication for GaP. Beginning with material losses for bulk GaP, we note that Fig. 2 in Ref. \cite{dean67a} considers the intrinsic optical absorption of GaP and finds an absorption coefficient $\alpha < 0.1/$cm for wavelengths larger than approximately $550$~nm, where the intensity attenuation is given by $I(z)=I_0e^{-\alpha z}$. These authors further explore low-level interband absorption due to Arsenic dopants \cite{dean69a}, emphasizing the need for high purity materials for our application.

More recently, nanophotonic structures have been fabricated in GaP and quality factors measured over a range of wavelengths in the visible and near infrared. Examples include a hybrid ring resonator of $900$~nm diameter with quality factor $Q \simeq 6800$ for $\lambda_0 \simeq 637$~nm \cite{barclay11a} and a photonic crystal cavity with lattice constant $d \simeq 200$~nm and measured $Q \simeq 12000$ for $\lambda_0$ around $800$~nm \cite{rivoire09a}. The associated absorption coefficients are $\alpha \simeq 24/$~cm and $\alpha \simeq 11/$cm, respectively, more than $10^2$ times larger than the limit from Ref. \cite{dean67a}.
The highest reported quality factor is $Q\simeq 2.8\times 10^5$ at $\lambda_0 \simeq 1.55$~$\mu$m for a GaP micro disk of diameter approximately $6$~$\mu$m \cite{mitchell14a}. The corresponding absorption coefficient $\alpha \simeq 0.24/$cm, which approaches the ultra-low loss regime of Ref. \cite{dean67a}.

These loss values for fabricated GaP structures represent steady progress of improved performance, including for a photonic crystal \cite{rivoire09a} comparable to those in our manuscript \cite{gonzaleztudelaMain}. For example, the structure in Fig. 3 \cite{gonzaleztudelaMain} has lattice constant $d=316$nm. A lattice of $60\times60$ unit cells would have linear dimension $L=60 d \simeq 20$~$\mu$m and could accommodate a few thousand atoms. For $\alpha \simeq 11/$cm as for the photonic crystal in Ref. \cite{rivoire09a}, the reduction in intensity for a propagating guided mode across the lattice would be $\alpha L \sim 2 \times 10^{-2}$ (i.e., $2\%$ loss) far from a band edge, with further reduction by perhaps $>10 \times$ suggested by the limits in Refs. \cite{dean67a,mitchell14a}. However, the loss would be further increased near a band edge, especially for the flat bands considered in our manuscript. We are currently investigating this issue by way of FDTD calculations for our proposed structures.
 
Apart from reducing absorption for guided modes used for atom-atom interactions, minimizing absorption in the PCW structures of our manuscript is critical since the intensities employed for atom trapping will be high. The largest Rabi frequency is $\Omega_{\mathrm{SI}}/2\pi = 130$GHz for the scheme in Fig. 2 \cite{gonzaleztudelaMain}, corresponding to intensity $I \sim 0.1$~W/$\mu$m$^2$. This high value is mitigated by the fact that the planar PCW is located near a node in the standing wave formed by the two counter propagating side illumination beams. Nevertheless, thermal management and optical nonlinearities of the dielectric will be important issues. 

To gain some perspective on these issues, we refer to the generation of frequency combs with microscopic ring resonators \cite{kippenberg11a}. Ref. \cite{levy10a} reports a Silicon Nitride (SiN) ring resonator of cross sectional area $\sim 3 \mu$m$^2$ pumped with $P_{in} \simeq 300$~mW leading to circulating power within the ring resonator $P_c \simeq 300$W and intracavity intensity $I \sim 10^{2}$~W/$\mu$m$^2$ without optical damage. The high intensity leads to nonlinear interactions to generate a frequency comb, from which the authors determine the value of the nonlinear index of refraction for SiN to be $n_2 \simeq 2.5\times10^{-15}$~cm$^2$/W (i.e., about $10\times$ larger than for silica).

For our structures with $I \sim 0.1$~W/$\mu$m$^2$ as in Fig. 2a for the vacuum lattice, the nonlinear phase shift across the thickness of the substrate $W \simeq 120$~nm would be $\delta \phi \lesssim 10^{-6}$~radians for a single side illumination beam. Here, we use the nonlinear index quoted in Ref. \cite{liu10a}, namely $n_2 \simeq 7\times10^{-14}$~cm$^2$/W determined for GaP illuminated with femtosecond laser pulses at $1040$nm.

Certainly, we recognize that these estimates provide only a basis for some optimism and motivation for further investigations of the feasibility of the structures described in our manuscript \cite{gonzaleztudelaMain}. The realization of our proposals for nanophotonic atomic lattices requires overcoming significant challenges in material characterization and device fabrication not by a literature survey but rather by a dedicated research program, including materials other than GaP.

%\section{Acknowledgements\label{sec:acknowledgements}}

\bibliography{Sci,books}

\begin{thebibliography}{60}
\bibitem[{foo()}]{footnote1}
\bibinfo{note}{We have neglected terms appearing with $a_k^\dagger |g_1\rangle
  \langle e|$ (and h.c.), as they will give no contribution to the master
  equation in the vaccuum regime, i.e., $\langle a_k^\dagger a_k \rangle=0$.}
\end{thebibliography}

\end{document}